\documentclass[aps,pre,onecolumn,superscriptaddress,floatfix]{revtex4}
\usepackage{epsf}
\usepackage{amsmath}
\usepackage{amssymb}
\usepackage{color}
\definecolor{AliceBlue}{rgb}{0.94,0.97,1.00}
\definecolor{AntiqueWhite1}{rgb}{1.00,0.94,0.86}
\definecolor{AntiqueWhite2}{rgb}{0.93,0.87,0.80}
\definecolor{AntiqueWhite3}{rgb}{0.80,0.75,0.69}
\definecolor{AntiqueWhite4}{rgb}{0.55,0.51,0.47}
\definecolor{AntiqueWhite}{rgb}{0.98,0.92,0.84}
\definecolor{BlanchedAlmond}{rgb}{1.00,0.92,0.80}
\definecolor{BlueViolet}{rgb}{0.54,0.17,0.89}
\definecolor{CadetBlue1}{rgb}{0.60,0.96,1.00}
\definecolor{CadetBlue2}{rgb}{0.56,0.90,0.93}
\definecolor{CadetBlue3}{rgb}{0.48,0.77,0.80}
\definecolor{CadetBlue4}{rgb}{0.33,0.53,0.55}
\definecolor{CadetBlue}{rgb}{0.37,0.62,0.63}
\definecolor{CornflowerBlue}{rgb}{0.39,0.58,0.93}
\definecolor{DarkBlue}{rgb}{0.00,0.00,0.55}
\definecolor{DarkCyan}{rgb}{0.00,0.55,0.55}
\definecolor{DarkGoldenrod1}{rgb}{1.00,0.73,0.06}
\definecolor{DarkGoldenrod2}{rgb}{0.93,0.68,0.05}
\definecolor{DarkGoldenrod3}{rgb}{0.80,0.58,0.05}
\definecolor{DarkGoldenrod4}{rgb}{0.55,0.40,0.03}
\definecolor{DarkGoldenrod}{rgb}{0.72,0.53,0.04}
\definecolor{DarkGray}{rgb}{0.66,0.66,0.66}
\definecolor{DarkGreen}{rgb}{0.00,0.39,0.00}
\definecolor{DarkGrey}{rgb}{0.66,0.66,0.66}
\definecolor{DarkKhaki}{rgb}{0.74,0.72,0.42}
\definecolor{DarkMagenta}{rgb}{0.55,0.00,0.55}
\definecolor{DarkOliveGreen1}{rgb}{0.79,1.00,0.44}
\definecolor{DarkOliveGreen2}{rgb}{0.74,0.93,0.41}
\definecolor{DarkOliveGreen3}{rgb}{0.64,0.80,0.35}
\definecolor{DarkOliveGreen4}{rgb}{0.43,0.55,0.24}
\definecolor{DarkOliveGreen}{rgb}{0.33,0.42,0.18}
\definecolor{DarkOrange1}{rgb}{1.00,0.50,0.00}
\definecolor{DarkOrange2}{rgb}{0.93,0.46,0.00}
\definecolor{DarkOrange3}{rgb}{0.80,0.40,0.00}
\definecolor{DarkOrange4}{rgb}{0.55,0.27,0.00}
\definecolor{DarkOrange}{rgb}{1.00,0.55,0.00}
\definecolor{DarkOrchid1}{rgb}{0.75,0.24,1.00}
\definecolor{DarkOrchid2}{rgb}{0.70,0.23,0.93}
\definecolor{DarkOrchid3}{rgb}{0.60,0.20,0.80}
\definecolor{DarkOrchid4}{rgb}{0.41,0.13,0.55}
\definecolor{DarkOrchid}{rgb}{0.60,0.20,0.80}
\definecolor{DarkRed}{rgb}{0.55,0.00,0.00}
\definecolor{DarkSalmon}{rgb}{0.91,0.59,0.48}
\definecolor{DarkSeaGreen1}{rgb}{0.76,1.00,0.76}
\definecolor{DarkSeaGreen2}{rgb}{0.71,0.93,0.71}
\definecolor{DarkSeaGreen3}{rgb}{0.61,0.80,0.61}
\definecolor{DarkSeaGreen4}{rgb}{0.41,0.55,0.41}
\definecolor{DarkSeaGreen}{rgb}{0.56,0.74,0.56}
\definecolor{DarkSlateBlue}{rgb}{0.28,0.24,0.55}
\definecolor{DarkSlateGray1}{rgb}{0.59,1.00,1.00}
\definecolor{DarkSlateGray2}{rgb}{0.55,0.93,0.93}
\definecolor{DarkSlateGray3}{rgb}{0.47,0.80,0.80}
\definecolor{DarkSlateGray4}{rgb}{0.32,0.55,0.55}
\definecolor{DarkSlateGray}{rgb}{0.18,0.31,0.31}
\definecolor{DarkSlateGrey}{rgb}{0.18,0.31,0.31}
\definecolor{DarkTurquoise}{rgb}{0.00,0.81,0.82}
\definecolor{DarkViolet}{rgb}{0.58,0.00,0.83}
\definecolor{DeepPink1}{rgb}{1.00,0.08,0.58}
\definecolor{DeepPink2}{rgb}{0.93,0.07,0.54}
\definecolor{DeepPink3}{rgb}{0.80,0.06,0.46}
\definecolor{DeepPink4}{rgb}{0.55,0.04,0.31}
\definecolor{DeepPink}{rgb}{1.00,0.08,0.58}
\definecolor{DeepSkyBlue1}{rgb}{0.00,0.75,1.00}
\definecolor{DeepSkyBlue2}{rgb}{0.00,0.70,0.93}
\definecolor{DeepSkyBlue3}{rgb}{0.00,0.60,0.80}
\definecolor{DeepSkyBlue4}{rgb}{0.00,0.41,0.55}
\definecolor{DeepSkyBlue}{rgb}{0.00,0.75,1.00}
\definecolor{DimGray}{rgb}{0.41,0.41,0.41}
\definecolor{DimGrey}{rgb}{0.41,0.41,0.41}
\definecolor{DodgerBlue1}{rgb}{0.12,0.56,1.00}
\definecolor{DodgerBlue2}{rgb}{0.11,0.53,0.93}
\definecolor{DodgerBlue3}{rgb}{0.09,0.45,0.80}
\definecolor{DodgerBlue4}{rgb}{0.06,0.31,0.55}
\definecolor{DodgerBlue}{rgb}{0.12,0.56,1.00}
\definecolor{FloralWhite}{rgb}{1.00,0.98,0.94}
\definecolor{ForestGreen}{rgb}{0.13,0.55,0.13}
\definecolor{GhostWhite}{rgb}{0.97,0.97,1.00}
\definecolor{GreenYellow}{rgb}{0.68,1.00,0.18}
\definecolor{HotPink1}{rgb}{1.00,0.43,0.71}
\definecolor{HotPink2}{rgb}{0.93,0.42,0.65}
\definecolor{HotPink3}{rgb}{0.80,0.38,0.56}
\definecolor{HotPink4}{rgb}{0.55,0.23,0.38}
\definecolor{HotPink}{rgb}{1.00,0.41,0.71}
\definecolor{IndianRed1}{rgb}{1.00,0.42,0.42}
\definecolor{IndianRed2}{rgb}{0.93,0.39,0.39}
\definecolor{IndianRed3}{rgb}{0.80,0.33,0.33}
\definecolor{IndianRed4}{rgb}{0.55,0.23,0.23}
\definecolor{IndianRed}{rgb}{0.80,0.36,0.36}
\definecolor{LavenderBlush1}{rgb}{1.00,0.94,0.96}
\definecolor{LavenderBlush2}{rgb}{0.93,0.88,0.90}
\definecolor{LavenderBlush3}{rgb}{0.80,0.76,0.77}
\definecolor{LavenderBlush4}{rgb}{0.55,0.51,0.53}
\definecolor{LavenderBlush}{rgb}{1.00,0.94,0.96}
\definecolor{LawnGreen}{rgb}{0.49,0.99,0.00}
\definecolor{LemonChiffon1}{rgb}{1.00,0.98,0.80}
\definecolor{LemonChiffon2}{rgb}{0.93,0.91,0.75}
\definecolor{LemonChiffon3}{rgb}{0.80,0.79,0.65}
\definecolor{LemonChiffon4}{rgb}{0.55,0.54,0.44}
\definecolor{LemonChiffon}{rgb}{1.00,0.98,0.80}
\definecolor{LightBlue1}{rgb}{0.75,0.94,1.00}
\definecolor{LightBlue2}{rgb}{0.70,0.87,0.93}
\definecolor{LightBlue3}{rgb}{0.60,0.75,0.80}
\definecolor{LightBlue4}{rgb}{0.41,0.51,0.55}
\definecolor{LightBlue}{rgb}{0.68,0.85,0.90}
\definecolor{LightCoral}{rgb}{0.94,0.50,0.50}
\definecolor{LightCyan1}{rgb}{0.88,1.00,1.00}
\definecolor{LightCyan2}{rgb}{0.82,0.93,0.93}
\definecolor{LightCyan3}{rgb}{0.71,0.80,0.80}
\definecolor{LightCyan4}{rgb}{0.48,0.55,0.55}
\definecolor{LightCyan}{rgb}{0.88,1.00,1.00}
\definecolor{LightGoldenrod1}{rgb}{1.00,0.93,0.55}
\definecolor{LightGoldenrod2}{rgb}{0.93,0.86,0.51}
\definecolor{LightGoldenrod3}{rgb}{0.80,0.75,0.44}
\definecolor{LightGoldenrod4}{rgb}{0.55,0.51,0.30}
\definecolor{LightGoldenrodYellow}{rgb}{0.98,0.98,0.82}
\definecolor{LightGoldenrod}{rgb}{0.93,0.87,0.51}
\definecolor{LightGray}{rgb}{0.83,0.83,0.83}
\definecolor{LightGreen}{rgb}{0.56,0.93,0.56}
\definecolor{LightGrey}{rgb}{0.83,0.83,0.83}
\definecolor{LightPink1}{rgb}{1.00,0.68,0.73}
\definecolor{LightPink2}{rgb}{0.93,0.64,0.68}
\definecolor{LightPink3}{rgb}{0.80,0.55,0.58}
\definecolor{LightPink4}{rgb}{0.55,0.37,0.40}
\definecolor{LightPink}{rgb}{1.00,0.71,0.76}
\definecolor{LightSalmon1}{rgb}{1.00,0.63,0.48}
\definecolor{LightSalmon2}{rgb}{0.93,0.58,0.45}
\definecolor{LightSalmon3}{rgb}{0.80,0.51,0.38}
\definecolor{LightSalmon4}{rgb}{0.55,0.34,0.26}
\definecolor{LightSalmon}{rgb}{1.00,0.63,0.48}
\definecolor{LightSeaGreen}{rgb}{0.13,0.70,0.67}
\definecolor{LightSkyBlue1}{rgb}{0.69,0.89,1.00}
\definecolor{LightSkyBlue2}{rgb}{0.64,0.83,0.93}
\definecolor{LightSkyBlue3}{rgb}{0.55,0.71,0.80}
\definecolor{LightSkyBlue4}{rgb}{0.38,0.48,0.55}
\definecolor{LightSkyBlue}{rgb}{0.53,0.81,0.98}
\definecolor{LightSlateBlue}{rgb}{0.52,0.44,1.00}
\definecolor{LightSlateGray}{rgb}{0.47,0.53,0.60}
\definecolor{LightSlateGrey}{rgb}{0.47,0.53,0.60}
\definecolor{LightSteelBlue1}{rgb}{0.79,0.88,1.00}
\definecolor{LightSteelBlue2}{rgb}{0.74,0.82,0.93}
\definecolor{LightSteelBlue3}{rgb}{0.64,0.71,0.80}
\definecolor{LightSteelBlue4}{rgb}{0.43,0.48,0.55}
\definecolor{LightSteelBlue}{rgb}{0.69,0.77,0.87}
\definecolor{LightYellow1}{rgb}{1.00,1.00,0.88}
\definecolor{LightYellow2}{rgb}{0.93,0.93,0.82}
\definecolor{LightYellow3}{rgb}{0.80,0.80,0.71}
\definecolor{LightYellow4}{rgb}{0.55,0.55,0.48}
\definecolor{LightYellow}{rgb}{1.00,1.00,0.88}
\definecolor{LimeGreen}{rgb}{0.20,0.80,0.20}
\definecolor{MediumAquamarine}{rgb}{0.40,0.80,0.67}
\definecolor{MediumBlue}{rgb}{0.00,0.00,0.80}
\definecolor{MediumOrchid1}{rgb}{0.88,0.40,1.00}
\definecolor{MediumOrchid2}{rgb}{0.82,0.37,0.93}
\definecolor{MediumOrchid3}{rgb}{0.71,0.32,0.80}
\definecolor{MediumOrchid4}{rgb}{0.48,0.22,0.55}
\definecolor{MediumOrchid}{rgb}{0.73,0.33,0.83}
\definecolor{MediumPurple1}{rgb}{0.67,0.51,1.00}
\definecolor{MediumPurple2}{rgb}{0.62,0.47,0.93}
\definecolor{MediumPurple3}{rgb}{0.54,0.41,0.80}
\definecolor{MediumPurple4}{rgb}{0.36,0.28,0.55}
\definecolor{MediumPurple}{rgb}{0.58,0.44,0.86}
\definecolor{MediumSeaGreen}{rgb}{0.24,0.70,0.44}
\definecolor{MediumSlateBlue}{rgb}{0.48,0.41,0.93}
\definecolor{MediumSpringGreen}{rgb}{0.00,0.98,0.60}
\definecolor{MediumTurquoise}{rgb}{0.28,0.82,0.80}
\definecolor{MediumVioletRed}{rgb}{0.78,0.08,0.52}
\definecolor{MidnightBlue}{rgb}{0.10,0.10,0.44}
\definecolor{MintCream}{rgb}{0.96,1.00,0.98}
\definecolor{MistyRose1}{rgb}{1.00,0.89,0.88}
\definecolor{MistyRose2}{rgb}{0.93,0.84,0.82}
\definecolor{MistyRose3}{rgb}{0.80,0.72,0.71}
\definecolor{MistyRose4}{rgb}{0.55,0.49,0.48}
\definecolor{MistyRose}{rgb}{1.00,0.89,0.88}
\definecolor{NavajoWhite1}{rgb}{1.00,0.87,0.68}
\definecolor{NavajoWhite2}{rgb}{0.93,0.81,0.63}
\definecolor{NavajoWhite3}{rgb}{0.80,0.70,0.55}
\definecolor{NavajoWhite4}{rgb}{0.55,0.47,0.37}
\definecolor{NavajoWhite}{rgb}{1.00,0.87,0.68}
\definecolor{NavyBlue}{rgb}{0.00,0.00,0.50}
\definecolor{OldLace}{rgb}{0.99,0.96,0.90}
\definecolor{OliveDrab1}{rgb}{0.75,1.00,0.24}
\definecolor{OliveDrab2}{rgb}{0.70,0.93,0.23}
\definecolor{OliveDrab3}{rgb}{0.60,0.80,0.20}
\definecolor{OliveDrab4}{rgb}{0.41,0.55,0.13}
\definecolor{OliveDrab}{rgb}{0.42,0.56,0.14}
\definecolor{OrangeRed1}{rgb}{1.00,0.27,0.00}
\definecolor{OrangeRed2}{rgb}{0.93,0.25,0.00}
\definecolor{OrangeRed3}{rgb}{0.80,0.22,0.00}
\definecolor{OrangeRed4}{rgb}{0.55,0.15,0.00}
\definecolor{OrangeRed}{rgb}{1.00,0.27,0.00}
\definecolor{PaleGoldenrod}{rgb}{0.93,0.91,0.67}
\definecolor{PaleGreen1}{rgb}{0.60,1.00,0.60}
\definecolor{PaleGreen2}{rgb}{0.56,0.93,0.56}
\definecolor{PaleGreen3}{rgb}{0.49,0.80,0.49}
\definecolor{PaleGreen4}{rgb}{0.33,0.55,0.33}
\definecolor{PaleGreen}{rgb}{0.60,0.98,0.60}
\definecolor{PaleTurquoise1}{rgb}{0.73,1.00,1.00}
\definecolor{PaleTurquoise2}{rgb}{0.68,0.93,0.93}
\definecolor{PaleTurquoise3}{rgb}{0.59,0.80,0.80}
\definecolor{PaleTurquoise4}{rgb}{0.40,0.55,0.55}
\definecolor{PaleTurquoise}{rgb}{0.69,0.93,0.93}
\definecolor{PaleVioletRed1}{rgb}{1.00,0.51,0.67}
\definecolor{PaleVioletRed2}{rgb}{0.93,0.47,0.62}
\definecolor{PaleVioletRed3}{rgb}{0.80,0.41,0.54}
\definecolor{PaleVioletRed4}{rgb}{0.55,0.28,0.36}
\definecolor{PaleVioletRed}{rgb}{0.86,0.44,0.58}
\definecolor{PapayaWhip}{rgb}{1.00,0.94,0.84}
\definecolor{PeachPuff1}{rgb}{1.00,0.85,0.73}
\definecolor{PeachPuff2}{rgb}{0.93,0.80,0.68}
\definecolor{PeachPuff3}{rgb}{0.80,0.69,0.58}
\definecolor{PeachPuff4}{rgb}{0.55,0.47,0.40}
\definecolor{PeachPuff}{rgb}{1.00,0.85,0.73}
\definecolor{PowderBlue}{rgb}{0.69,0.88,0.90}
\definecolor{RosyBrown1}{rgb}{1.00,0.76,0.76}
\definecolor{RosyBrown2}{rgb}{0.93,0.71,0.71}
\definecolor{RosyBrown3}{rgb}{0.80,0.61,0.61}
\definecolor{RosyBrown4}{rgb}{0.55,0.41,0.41}
\definecolor{RosyBrown}{rgb}{0.74,0.56,0.56}
\definecolor{RoyalBlue1}{rgb}{0.28,0.46,1.00}
\definecolor{RoyalBlue2}{rgb}{0.26,0.43,0.93}
\definecolor{RoyalBlue3}{rgb}{0.23,0.37,0.80}
\definecolor{RoyalBlue4}{rgb}{0.15,0.25,0.55}
\definecolor{RoyalBlue}{rgb}{0.25,0.41,0.88}
\definecolor{SaddleBrown}{rgb}{0.55,0.27,0.07}
\definecolor{SandyBrown}{rgb}{0.96,0.64,0.38}
\definecolor{SeaGreen1}{rgb}{0.33,1.00,0.62}
\definecolor{SeaGreen2}{rgb}{0.31,0.93,0.58}
\definecolor{SeaGreen3}{rgb}{0.26,0.80,0.50}
\definecolor{SeaGreen4}{rgb}{0.18,0.55,0.34}
\definecolor{SeaGreen}{rgb}{0.18,0.55,0.34}
\definecolor{SkyBlue1}{rgb}{0.53,0.81,1.00}
\definecolor{SkyBlue2}{rgb}{0.49,0.75,0.93}
\definecolor{SkyBlue3}{rgb}{0.42,0.65,0.80}
\definecolor{SkyBlue4}{rgb}{0.29,0.44,0.55}
\definecolor{SkyBlue}{rgb}{0.53,0.81,0.92}
\definecolor{SlateBlue1}{rgb}{0.51,0.44,1.00}
\definecolor{SlateBlue2}{rgb}{0.48,0.40,0.93}
\definecolor{SlateBlue3}{rgb}{0.41,0.35,0.80}
\definecolor{SlateBlue4}{rgb}{0.28,0.24,0.55}
\definecolor{SlateBlue}{rgb}{0.42,0.35,0.80}
\definecolor{SlateGray1}{rgb}{0.78,0.89,1.00}
\definecolor{SlateGray2}{rgb}{0.73,0.83,0.93}
\definecolor{SlateGray3}{rgb}{0.62,0.71,0.80}
\definecolor{SlateGray4}{rgb}{0.42,0.48,0.55}
\definecolor{SlateGray}{rgb}{0.44,0.50,0.56}
\definecolor{SlateGrey}{rgb}{0.44,0.50,0.56}
\definecolor{SpringGreen1}{rgb}{0.00,1.00,0.50}
\definecolor{SpringGreen2}{rgb}{0.00,0.93,0.46}
\definecolor{SpringGreen3}{rgb}{0.00,0.80,0.40}
\definecolor{SpringGreen4}{rgb}{0.00,0.55,0.27}
\definecolor{SpringGreen}{rgb}{0.00,1.00,0.50}
\definecolor{SteelBlue1}{rgb}{0.39,0.72,1.00}
\definecolor{SteelBlue2}{rgb}{0.36,0.67,0.93}
\definecolor{SteelBlue3}{rgb}{0.31,0.58,0.80}
\definecolor{SteelBlue4}{rgb}{0.21,0.39,0.55}
\definecolor{SteelBlue}{rgb}{0.27,0.51,0.71}
\definecolor{VioletRed1}{rgb}{1.00,0.24,0.59}
\definecolor{VioletRed2}{rgb}{0.93,0.23,0.55}
\definecolor{VioletRed3}{rgb}{0.80,0.20,0.47}
\definecolor{VioletRed4}{rgb}{0.55,0.13,0.32}
\definecolor{VioletRed}{rgb}{0.82,0.13,0.56}
\definecolor{WhiteSmoke}{rgb}{0.96,0.96,0.96}
\definecolor{YellowGreen}{rgb}{0.60,0.80,0.20}
\definecolor{aliceblue}{rgb}{0.94,0.97,1.00}
\definecolor{antiquewhite}{rgb}{0.98,0.92,0.84}
\definecolor{aquamarine1}{rgb}{0.50,1.00,0.83}
\definecolor{aquamarine2}{rgb}{0.46,0.93,0.78}
\definecolor{aquamarine3}{rgb}{0.40,0.80,0.67}
\definecolor{aquamarine4}{rgb}{0.27,0.55,0.45}
\definecolor{aquamarine}{rgb}{0.50,1.00,0.83}
\definecolor{azure1}{rgb}{0.94,1.00,1.00}
\definecolor{azure2}{rgb}{0.88,0.93,0.93}
\definecolor{azure3}{rgb}{0.76,0.80,0.80}
\definecolor{azure4}{rgb}{0.51,0.55,0.55}
\definecolor{azure}{rgb}{0.94,1.00,1.00}
\definecolor{beige}{rgb}{0.96,0.96,0.86}
\definecolor{bisque1}{rgb}{1.00,0.89,0.77}
\definecolor{bisque2}{rgb}{0.93,0.84,0.72}
\definecolor{bisque3}{rgb}{0.80,0.72,0.62}
\definecolor{bisque4}{rgb}{0.55,0.49,0.42}
\definecolor{bisque}{rgb}{1.00,0.89,0.77}
\definecolor{black}{rgb}{0.00,0.00,0.00}
\definecolor{blanchedalmond}{rgb}{1.00,0.92,0.80}
\definecolor{blue1}{rgb}{0.00,0.00,1.00}
\definecolor{blue2}{rgb}{0.00,0.00,0.93}
\definecolor{blue3}{rgb}{0.00,0.00,0.80}
\definecolor{blue4}{rgb}{0.00,0.00,0.55}
\definecolor{blueviolet}{rgb}{0.54,0.17,0.89}
\definecolor{blue}{rgb}{0.00,0.00,1.00}
\definecolor{brown1}{rgb}{1.00,0.25,0.25}
\definecolor{brown2}{rgb}{0.93,0.23,0.23}
\definecolor{brown3}{rgb}{0.80,0.20,0.20}
\definecolor{brown4}{rgb}{0.55,0.14,0.14}
\definecolor{brown}{rgb}{0.65,0.16,0.16}
\definecolor{burlywood1}{rgb}{1.00,0.83,0.61}
\definecolor{burlywood2}{rgb}{0.93,0.77,0.57}
\definecolor{burlywood3}{rgb}{0.80,0.67,0.49}
\definecolor{burlywood4}{rgb}{0.55,0.45,0.33}
\definecolor{burlywood}{rgb}{0.87,0.72,0.53}
\definecolor{cadetblue}{rgb}{0.37,0.62,0.63}
\definecolor{chartreuse1}{rgb}{0.50,1.00,0.00}
\definecolor{chartreuse2}{rgb}{0.46,0.93,0.00}
\definecolor{chartreuse3}{rgb}{0.40,0.80,0.00}
\definecolor{chartreuse4}{rgb}{0.27,0.55,0.00}
\definecolor{chartreuse}{rgb}{0.50,1.00,0.00}
\definecolor{chocolate1}{rgb}{1.00,0.50,0.14}
\definecolor{chocolate2}{rgb}{0.93,0.46,0.13}
\definecolor{chocolate3}{rgb}{0.80,0.40,0.11}
\definecolor{chocolate4}{rgb}{0.55,0.27,0.07}
\definecolor{chocolate}{rgb}{0.82,0.41,0.12}
\definecolor{coral1}{rgb}{1.00,0.45,0.34}
\definecolor{coral2}{rgb}{0.93,0.42,0.31}
\definecolor{coral3}{rgb}{0.80,0.36,0.27}
\definecolor{coral4}{rgb}{0.55,0.24,0.18}
\definecolor{coral}{rgb}{1.00,0.50,0.31}
\definecolor{cornflowerblue}{rgb}{0.39,0.58,0.93}
\definecolor{cornsilk1}{rgb}{1.00,0.97,0.86}
\definecolor{cornsilk2}{rgb}{0.93,0.91,0.80}
\definecolor{cornsilk3}{rgb}{0.80,0.78,0.69}
\definecolor{cornsilk4}{rgb}{0.55,0.53,0.47}
\definecolor{cornsilk}{rgb}{1.00,0.97,0.86}
\definecolor{cyan1}{rgb}{0.00,1.00,1.00}
\definecolor{cyan2}{rgb}{0.00,0.93,0.93}
\definecolor{cyan3}{rgb}{0.00,0.80,0.80}
\definecolor{cyan4}{rgb}{0.00,0.55,0.55}
\definecolor{cyan}{rgb}{0.00,1.00,1.00}
\definecolor{darkblue}{rgb}{0.00,0.00,0.55}
\definecolor{darkcyan}{rgb}{0.00,0.55,0.55}
\definecolor{darkgoldenrod}{rgb}{0.72,0.53,0.04}
\definecolor{darkgray}{rgb}{0.66,0.66,0.66}
\definecolor{darkgreen}{rgb}{0.00,0.39,0.00}
\definecolor{darkgrey}{rgb}{0.66,0.66,0.66}
\definecolor{darkkhaki}{rgb}{0.74,0.72,0.42}
\definecolor{darkmagenta}{rgb}{0.55,0.00,0.55}
\definecolor{darkolive}{rgb}{0.33,0.42,0.18}
\definecolor{darkorange}{rgb}{1.00,0.55,0.00}
\definecolor{darkorchid}{rgb}{0.60,0.20,0.80}
\definecolor{darkred}{rgb}{0.55,0.00,0.00}
\definecolor{darksalmon}{rgb}{0.91,0.59,0.48}
\definecolor{darksea}{rgb}{0.56,0.74,0.56}
\definecolor{darkslate}{rgb}{0.18,0.31,0.31}
\definecolor{darkslate}{rgb}{0.18,0.31,0.31}
\definecolor{darkslate}{rgb}{0.28,0.24,0.55}
\definecolor{darkturquoise}{rgb}{0.00,0.81,0.82}
\definecolor{darkviolet}{rgb}{0.58,0.00,0.83}
\definecolor{deeppink}{rgb}{1.00,0.08,0.58}
\definecolor{deepsky}{rgb}{0.00,0.75,1.00}
\definecolor{dimgray}{rgb}{0.41,0.41,0.41}
\definecolor{dimgrey}{rgb}{0.41,0.41,0.41}
\definecolor{dodgerblue}{rgb}{0.12,0.56,1.00}
\definecolor{firebrick1}{rgb}{1.00,0.19,0.19}
\definecolor{firebrick2}{rgb}{0.93,0.17,0.17}
\definecolor{firebrick3}{rgb}{0.80,0.15,0.15}
\definecolor{firebrick4}{rgb}{0.55,0.10,0.10}
\definecolor{firebrick}{rgb}{0.70,0.13,0.13}
\definecolor{floralwhite}{rgb}{1.00,0.98,0.94}
\definecolor{forestgreen}{rgb}{0.13,0.55,0.13}
\definecolor{gainsboro}{rgb}{0.86,0.86,0.86}
\definecolor{ghostwhite}{rgb}{0.97,0.97,1.00}
\definecolor{gold1}{rgb}{1.00,0.84,0.00}
\definecolor{gold2}{rgb}{0.93,0.79,0.00}
\definecolor{gold3}{rgb}{0.80,0.68,0.00}
\definecolor{gold4}{rgb}{0.55,0.46,0.00}
\definecolor{goldenrod1}{rgb}{1.00,0.76,0.15}
\definecolor{goldenrod2}{rgb}{0.93,0.71,0.13}
\definecolor{goldenrod3}{rgb}{0.80,0.61,0.11}
\definecolor{goldenrod4}{rgb}{0.55,0.41,0.08}
\definecolor{goldenrod}{rgb}{0.85,0.65,0.13}
\definecolor{gold}{rgb}{1.00,0.84,0.00}
\definecolor{gray0}{rgb}{0.00,0.00,0.00}
\definecolor{gray100}{rgb}{1.00,1.00,1.00}
\definecolor{gray10}{rgb}{0.10,0.10,0.10}
\definecolor{gray11}{rgb}{0.11,0.11,0.11}
\definecolor{gray12}{rgb}{0.12,0.12,0.12}
\definecolor{gray13}{rgb}{0.13,0.13,0.13}
\definecolor{gray14}{rgb}{0.14,0.14,0.14}
\definecolor{gray15}{rgb}{0.15,0.15,0.15}
\definecolor{gray16}{rgb}{0.16,0.16,0.16}
\definecolor{gray17}{rgb}{0.17,0.17,0.17}
\definecolor{gray18}{rgb}{0.18,0.18,0.18}
\definecolor{gray19}{rgb}{0.19,0.19,0.19}
\definecolor{gray1}{rgb}{0.01,0.01,0.01}
\definecolor{gray20}{rgb}{0.20,0.20,0.20}
\definecolor{gray21}{rgb}{0.21,0.21,0.21}
\definecolor{gray22}{rgb}{0.22,0.22,0.22}
\definecolor{gray23}{rgb}{0.23,0.23,0.23}
\definecolor{gray24}{rgb}{0.24,0.24,0.24}
\definecolor{gray25}{rgb}{0.25,0.25,0.25}
\definecolor{gray26}{rgb}{0.26,0.26,0.26}
\definecolor{gray27}{rgb}{0.27,0.27,0.27}
\definecolor{gray28}{rgb}{0.28,0.28,0.28}
\definecolor{gray29}{rgb}{0.29,0.29,0.29}
\definecolor{gray2}{rgb}{0.02,0.02,0.02}
\definecolor{gray30}{rgb}{0.30,0.30,0.30}
\definecolor{gray31}{rgb}{0.31,0.31,0.31}
\definecolor{gray32}{rgb}{0.32,0.32,0.32}
\definecolor{gray33}{rgb}{0.33,0.33,0.33}
\definecolor{gray34}{rgb}{0.34,0.34,0.34}
\definecolor{gray35}{rgb}{0.35,0.35,0.35}
\definecolor{gray36}{rgb}{0.36,0.36,0.36}
\definecolor{gray37}{rgb}{0.37,0.37,0.37}
\definecolor{gray38}{rgb}{0.38,0.38,0.38}
\definecolor{gray39}{rgb}{0.39,0.39,0.39}
\definecolor{gray3}{rgb}{0.03,0.03,0.03}
\definecolor{gray40}{rgb}{0.40,0.40,0.40}
\definecolor{gray41}{rgb}{0.41,0.41,0.41}
\definecolor{gray42}{rgb}{0.42,0.42,0.42}
\definecolor{gray43}{rgb}{0.43,0.43,0.43}
\definecolor{gray44}{rgb}{0.44,0.44,0.44}
\definecolor{gray45}{rgb}{0.45,0.45,0.45}
\definecolor{gray46}{rgb}{0.46,0.46,0.46}
\definecolor{gray47}{rgb}{0.47,0.47,0.47}
\definecolor{gray48}{rgb}{0.48,0.48,0.48}
\definecolor{gray49}{rgb}{0.49,0.49,0.49}
\definecolor{gray4}{rgb}{0.04,0.04,0.04}
\definecolor{gray50}{rgb}{0.50,0.50,0.50}
\definecolor{gray51}{rgb}{0.51,0.51,0.51}
\definecolor{gray52}{rgb}{0.52,0.52,0.52}
\definecolor{gray53}{rgb}{0.53,0.53,0.53}
\definecolor{gray54}{rgb}{0.54,0.54,0.54}
\definecolor{gray55}{rgb}{0.55,0.55,0.55}
\definecolor{gray56}{rgb}{0.56,0.56,0.56}
\definecolor{gray57}{rgb}{0.57,0.57,0.57}
\definecolor{gray58}{rgb}{0.58,0.58,0.58}
\definecolor{gray59}{rgb}{0.59,0.59,0.59}
\definecolor{gray5}{rgb}{0.05,0.05,0.05}
\definecolor{gray60}{rgb}{0.60,0.60,0.60}
\definecolor{gray61}{rgb}{0.61,0.61,0.61}
\definecolor{gray62}{rgb}{0.62,0.62,0.62}
\definecolor{gray63}{rgb}{0.63,0.63,0.63}
\definecolor{gray64}{rgb}{0.64,0.64,0.64}
\definecolor{gray65}{rgb}{0.65,0.65,0.65}
\definecolor{gray66}{rgb}{0.66,0.66,0.66}
\definecolor{gray67}{rgb}{0.67,0.67,0.67}
\definecolor{gray68}{rgb}{0.68,0.68,0.68}
\definecolor{gray69}{rgb}{0.69,0.69,0.69}
\definecolor{gray6}{rgb}{0.06,0.06,0.06}
\definecolor{gray70}{rgb}{0.70,0.70,0.70}
\definecolor{gray71}{rgb}{0.71,0.71,0.71}
\definecolor{gray72}{rgb}{0.72,0.72,0.72}
\definecolor{gray73}{rgb}{0.73,0.73,0.73}
\definecolor{gray74}{rgb}{0.74,0.74,0.74}
\definecolor{gray75}{rgb}{0.75,0.75,0.75}
\definecolor{gray76}{rgb}{0.76,0.76,0.76}
\definecolor{gray77}{rgb}{0.77,0.77,0.77}
\definecolor{gray78}{rgb}{0.78,0.78,0.78}
\definecolor{gray79}{rgb}{0.79,0.79,0.79}
\definecolor{gray7}{rgb}{0.07,0.07,0.07}
\definecolor{gray80}{rgb}{0.80,0.80,0.80}
\definecolor{gray81}{rgb}{0.81,0.81,0.81}
\definecolor{gray82}{rgb}{0.82,0.82,0.82}
\definecolor{gray83}{rgb}{0.83,0.83,0.83}
\definecolor{gray84}{rgb}{0.84,0.84,0.84}
\definecolor{gray85}{rgb}{0.85,0.85,0.85}
\definecolor{gray86}{rgb}{0.86,0.86,0.86}
\definecolor{gray87}{rgb}{0.87,0.87,0.87}
\definecolor{gray88}{rgb}{0.88,0.88,0.88}
\definecolor{gray89}{rgb}{0.89,0.89,0.89}
\definecolor{gray8}{rgb}{0.08,0.08,0.08}
\definecolor{gray90}{rgb}{0.90,0.90,0.90}
\definecolor{gray91}{rgb}{0.91,0.91,0.91}
\definecolor{gray92}{rgb}{0.92,0.92,0.92}
\definecolor{gray93}{rgb}{0.93,0.93,0.93}
\definecolor{gray94}{rgb}{0.94,0.94,0.94}
\definecolor{gray95}{rgb}{0.95,0.95,0.95}
\definecolor{gray96}{rgb}{0.96,0.96,0.96}
\definecolor{gray97}{rgb}{0.97,0.97,0.97}
\definecolor{gray98}{rgb}{0.98,0.98,0.98}
\definecolor{gray99}{rgb}{0.99,0.99,0.99}
\definecolor{gray9}{rgb}{0.09,0.09,0.09}
\definecolor{gray}{rgb}{0.75,0.75,0.75}
\definecolor{green1}{rgb}{0.00,1.00,0.00}
\definecolor{green2}{rgb}{0.00,0.93,0.00}
\definecolor{green3}{rgb}{0.00,0.80,0.00}
\definecolor{green4}{rgb}{0.00,0.55,0.00}
\definecolor{greenyellow}{rgb}{0.68,1.00,0.18}
\definecolor{green}{rgb}{0.00,1.00,0.00}
\definecolor{grey0}{rgb}{0.00,0.00,0.00}
\definecolor{grey100}{rgb}{1.00,1.00,1.00}
\definecolor{grey10}{rgb}{0.10,0.10,0.10}
\definecolor{grey11}{rgb}{0.11,0.11,0.11}
\definecolor{grey12}{rgb}{0.12,0.12,0.12}
\definecolor{grey13}{rgb}{0.13,0.13,0.13}
\definecolor{grey14}{rgb}{0.14,0.14,0.14}
\definecolor{grey15}{rgb}{0.15,0.15,0.15}
\definecolor{grey16}{rgb}{0.16,0.16,0.16}
\definecolor{grey17}{rgb}{0.17,0.17,0.17}
\definecolor{grey18}{rgb}{0.18,0.18,0.18}
\definecolor{grey19}{rgb}{0.19,0.19,0.19}
\definecolor{grey1}{rgb}{0.01,0.01,0.01}
\definecolor{grey20}{rgb}{0.20,0.20,0.20}
\definecolor{grey21}{rgb}{0.21,0.21,0.21}
\definecolor{grey22}{rgb}{0.22,0.22,0.22}
\definecolor{grey23}{rgb}{0.23,0.23,0.23}
\definecolor{grey24}{rgb}{0.24,0.24,0.24}
\definecolor{grey25}{rgb}{0.25,0.25,0.25}
\definecolor{grey26}{rgb}{0.26,0.26,0.26}
\definecolor{grey27}{rgb}{0.27,0.27,0.27}
\definecolor{grey28}{rgb}{0.28,0.28,0.28}
\definecolor{grey29}{rgb}{0.29,0.29,0.29}
\definecolor{grey2}{rgb}{0.02,0.02,0.02}
\definecolor{grey30}{rgb}{0.30,0.30,0.30}
\definecolor{grey31}{rgb}{0.31,0.31,0.31}
\definecolor{grey32}{rgb}{0.32,0.32,0.32}
\definecolor{grey33}{rgb}{0.33,0.33,0.33}
\definecolor{grey34}{rgb}{0.34,0.34,0.34}
\definecolor{grey35}{rgb}{0.35,0.35,0.35}
\definecolor{grey36}{rgb}{0.36,0.36,0.36}
\definecolor{grey37}{rgb}{0.37,0.37,0.37}
\definecolor{grey38}{rgb}{0.38,0.38,0.38}
\definecolor{grey39}{rgb}{0.39,0.39,0.39}
\definecolor{grey3}{rgb}{0.03,0.03,0.03}
\definecolor{grey40}{rgb}{0.40,0.40,0.40}
\definecolor{grey41}{rgb}{0.41,0.41,0.41}
\definecolor{grey42}{rgb}{0.42,0.42,0.42}
\definecolor{grey43}{rgb}{0.43,0.43,0.43}
\definecolor{grey44}{rgb}{0.44,0.44,0.44}
\definecolor{grey45}{rgb}{0.45,0.45,0.45}
\definecolor{grey46}{rgb}{0.46,0.46,0.46}
\definecolor{grey47}{rgb}{0.47,0.47,0.47}
\definecolor{grey48}{rgb}{0.48,0.48,0.48}
\definecolor{grey49}{rgb}{0.49,0.49,0.49}
\definecolor{grey4}{rgb}{0.04,0.04,0.04}
\definecolor{grey50}{rgb}{0.50,0.50,0.50}
\definecolor{grey51}{rgb}{0.51,0.51,0.51}
\definecolor{grey52}{rgb}{0.52,0.52,0.52}
\definecolor{grey53}{rgb}{0.53,0.53,0.53}
\definecolor{grey54}{rgb}{0.54,0.54,0.54}
\definecolor{grey55}{rgb}{0.55,0.55,0.55}
\definecolor{grey56}{rgb}{0.56,0.56,0.56}
\definecolor{grey57}{rgb}{0.57,0.57,0.57}
\definecolor{grey58}{rgb}{0.58,0.58,0.58}
\definecolor{grey59}{rgb}{0.59,0.59,0.59}
\definecolor{grey5}{rgb}{0.05,0.05,0.05}
\definecolor{grey60}{rgb}{0.60,0.60,0.60}
\definecolor{grey61}{rgb}{0.61,0.61,0.61}
\definecolor{grey62}{rgb}{0.62,0.62,0.62}
\definecolor{grey63}{rgb}{0.63,0.63,0.63}
\definecolor{grey64}{rgb}{0.64,0.64,0.64}
\definecolor{grey65}{rgb}{0.65,0.65,0.65}
\definecolor{grey66}{rgb}{0.66,0.66,0.66}
\definecolor{grey67}{rgb}{0.67,0.67,0.67}
\definecolor{grey68}{rgb}{0.68,0.68,0.68}
\definecolor{grey69}{rgb}{0.69,0.69,0.69}
\definecolor{grey6}{rgb}{0.06,0.06,0.06}
\definecolor{grey70}{rgb}{0.70,0.70,0.70}
\definecolor{grey71}{rgb}{0.71,0.71,0.71}
\definecolor{grey72}{rgb}{0.72,0.72,0.72}
\definecolor{grey73}{rgb}{0.73,0.73,0.73}
\definecolor{grey74}{rgb}{0.74,0.74,0.74}
\definecolor{grey75}{rgb}{0.75,0.75,0.75}
\definecolor{grey76}{rgb}{0.76,0.76,0.76}
\definecolor{grey77}{rgb}{0.77,0.77,0.77}
\definecolor{grey78}{rgb}{0.78,0.78,0.78}
\definecolor{grey79}{rgb}{0.79,0.79,0.79}
\definecolor{grey7}{rgb}{0.07,0.07,0.07}
\definecolor{grey80}{rgb}{0.80,0.80,0.80}
\definecolor{grey81}{rgb}{0.81,0.81,0.81}
\definecolor{grey82}{rgb}{0.82,0.82,0.82}
\definecolor{grey83}{rgb}{0.83,0.83,0.83}
\definecolor{grey84}{rgb}{0.84,0.84,0.84}
\definecolor{grey85}{rgb}{0.85,0.85,0.85}
\definecolor{grey86}{rgb}{0.86,0.86,0.86}
\definecolor{grey87}{rgb}{0.87,0.87,0.87}
\definecolor{grey88}{rgb}{0.88,0.88,0.88}
\definecolor{grey89}{rgb}{0.89,0.89,0.89}
\definecolor{grey8}{rgb}{0.08,0.08,0.08}
\definecolor{grey90}{rgb}{0.90,0.90,0.90}
\definecolor{grey91}{rgb}{0.91,0.91,0.91}
\definecolor{grey92}{rgb}{0.92,0.92,0.92}
\definecolor{grey93}{rgb}{0.93,0.93,0.93}
\definecolor{grey94}{rgb}{0.94,0.94,0.94}
\definecolor{grey95}{rgb}{0.95,0.95,0.95}
\definecolor{grey96}{rgb}{0.96,0.96,0.96}
\definecolor{grey97}{rgb}{0.97,0.97,0.97}
\definecolor{grey98}{rgb}{0.98,0.98,0.98}
\definecolor{grey99}{rgb}{0.99,0.99,0.99}
\definecolor{grey9}{rgb}{0.09,0.09,0.09}
\definecolor{grey}{rgb}{0.75,0.75,0.75}
\definecolor{honeydew1}{rgb}{0.94,1.00,0.94}
\definecolor{honeydew2}{rgb}{0.88,0.93,0.88}
\definecolor{honeydew3}{rgb}{0.76,0.80,0.76}
\definecolor{honeydew4}{rgb}{0.51,0.55,0.51}
\definecolor{honeydew}{rgb}{0.94,1.00,0.94}
\definecolor{hotpink}{rgb}{1.00,0.41,0.71}
\definecolor{indianred}{rgb}{0.80,0.36,0.36}
\definecolor{ivory1}{rgb}{1.00,1.00,0.94}
\definecolor{ivory2}{rgb}{0.93,0.93,0.88}
\definecolor{ivory3}{rgb}{0.80,0.80,0.76}
\definecolor{ivory4}{rgb}{0.55,0.55,0.51}
\definecolor{ivory}{rgb}{1.00,1.00,0.94}
\definecolor{khaki1}{rgb}{1.00,0.96,0.56}
\definecolor{khaki2}{rgb}{0.93,0.90,0.52}
\definecolor{khaki3}{rgb}{0.80,0.78,0.45}
\definecolor{khaki4}{rgb}{0.55,0.53,0.31}
\definecolor{khaki}{rgb}{0.94,0.90,0.55}
\definecolor{lavenderblush}{rgb}{1.00,0.94,0.96}
\definecolor{lavender}{rgb}{0.90,0.90,0.98}
\definecolor{lawngreen}{rgb}{0.49,0.99,0.00}
\definecolor{lemonchiffon}{rgb}{1.00,0.98,0.80}
\definecolor{lightblue}{rgb}{0.68,0.85,0.90}
\definecolor{lightcoral}{rgb}{0.94,0.50,0.50}
\definecolor{lightcyan}{rgb}{0.88,1.00,1.00}
\definecolor{lightgoldenrod}{rgb}{0.93,0.87,0.51}
\definecolor{lightgoldenrod}{rgb}{0.98,0.98,0.82}
\definecolor{lightgray}{rgb}{0.83,0.83,0.83}
\definecolor{lightgreen}{rgb}{0.56,0.93,0.56}
\definecolor{lightgrey}{rgb}{0.83,0.83,0.83}
\definecolor{lightpink}{rgb}{1.00,0.71,0.76}
\definecolor{lightsalmon}{rgb}{1.00,0.63,0.48}
\definecolor{lightsea}{rgb}{0.13,0.70,0.67}
\definecolor{lightsky}{rgb}{0.53,0.81,0.98}
\definecolor{lightslate}{rgb}{0.47,0.53,0.60}
\definecolor{lightslate}{rgb}{0.47,0.53,0.60}
\definecolor{lightslate}{rgb}{0.52,0.44,1.00}
\definecolor{lightsteel}{rgb}{0.69,0.77,0.87}
\definecolor{lightyellow}{rgb}{1.00,1.00,0.88}
\definecolor{limegreen}{rgb}{0.20,0.80,0.20}
\definecolor{linen}{rgb}{0.98,0.94,0.90}
\definecolor{magenta1}{rgb}{1.00,0.00,1.00}
\definecolor{magenta2}{rgb}{0.93,0.00,0.93}
\definecolor{magenta3}{rgb}{0.80,0.00,0.80}
\definecolor{magenta4}{rgb}{0.55,0.00,0.55}
\definecolor{magenta}{rgb}{1.00,0.00,1.00}
\definecolor{maroon1}{rgb}{1.00,0.20,0.70}
\definecolor{maroon2}{rgb}{0.93,0.19,0.65}
\definecolor{maroon3}{rgb}{0.80,0.16,0.56}
\definecolor{maroon4}{rgb}{0.55,0.11,0.38}
\definecolor{maroon}{rgb}{0.69,0.19,0.38}
\definecolor{mediumaquamarine}{rgb}{0.40,0.80,0.67}
\definecolor{mediumblue}{rgb}{0.00,0.00,0.80}
\definecolor{mediumorchid}{rgb}{0.73,0.33,0.83}
\definecolor{mediumpurple}{rgb}{0.58,0.44,0.86}
\definecolor{mediumsea}{rgb}{0.24,0.70,0.44}
\definecolor{mediumslate}{rgb}{0.48,0.41,0.93}
\definecolor{mediumspring}{rgb}{0.00,0.98,0.60}
\definecolor{mediumturquoise}{rgb}{0.28,0.82,0.80}
\definecolor{mediumviolet}{rgb}{0.78,0.08,0.52}
\definecolor{midnightblue}{rgb}{0.10,0.10,0.44}
\definecolor{mintcream}{rgb}{0.96,1.00,0.98}
\definecolor{mistyrose}{rgb}{1.00,0.89,0.88}
\definecolor{moccasin}{rgb}{1.00,0.89,0.71}
\definecolor{navajowhite}{rgb}{1.00,0.87,0.68}
\definecolor{navyblue}{rgb}{0.00,0.00,0.50}
\definecolor{navy}{rgb}{0.00,0.00,0.50}
\definecolor{oldlace}{rgb}{0.99,0.96,0.90}
\definecolor{olivedrab}{rgb}{0.42,0.56,0.14}
\definecolor{orange1}{rgb}{1.00,0.65,0.00}
\definecolor{orange2}{rgb}{0.93,0.60,0.00}
\definecolor{orange3}{rgb}{0.80,0.52,0.00}
\definecolor{orange4}{rgb}{0.55,0.35,0.00}
\definecolor{orangered}{rgb}{1.00,0.27,0.00}
\definecolor{orange}{rgb}{1.00,0.65,0.00}
\definecolor{orchid1}{rgb}{1.00,0.51,0.98}
\definecolor{orchid2}{rgb}{0.93,0.48,0.91}
\definecolor{orchid3}{rgb}{0.80,0.41,0.79}
\definecolor{orchid4}{rgb}{0.55,0.28,0.54}
\definecolor{orchid}{rgb}{0.85,0.44,0.84}
\definecolor{palegoldenrod}{rgb}{0.93,0.91,0.67}
\definecolor{palegreen}{rgb}{0.60,0.98,0.60}
\definecolor{paleturquoise}{rgb}{0.69,0.93,0.93}
\definecolor{paleviolet}{rgb}{0.86,0.44,0.58}
\definecolor{papayawhip}{rgb}{1.00,0.94,0.84}
\definecolor{peachpuff}{rgb}{1.00,0.85,0.73}
\definecolor{peru}{rgb}{0.80,0.52,0.25}
\definecolor{pink1}{rgb}{1.00,0.71,0.77}
\definecolor{pink2}{rgb}{0.93,0.66,0.72}
\definecolor{pink3}{rgb}{0.80,0.57,0.62}
\definecolor{pink4}{rgb}{0.55,0.39,0.42}
\definecolor{pink}{rgb}{1.00,0.75,0.80}
\definecolor{plum1}{rgb}{1.00,0.73,1.00}
\definecolor{plum2}{rgb}{0.93,0.68,0.93}
\definecolor{plum3}{rgb}{0.80,0.59,0.80}
\definecolor{plum4}{rgb}{0.55,0.40,0.55}
\definecolor{plum}{rgb}{0.87,0.63,0.87}
\definecolor{powderblue}{rgb}{0.69,0.88,0.90}
\definecolor{purple1}{rgb}{0.61,0.19,1.00}
\definecolor{purple2}{rgb}{0.57,0.17,0.93}
\definecolor{purple3}{rgb}{0.49,0.15,0.80}
\definecolor{purple4}{rgb}{0.33,0.10,0.55}
\definecolor{purple}{rgb}{0.63,0.13,0.94}
\definecolor{red1}{rgb}{1.00,0.00,0.00}
\definecolor{red2}{rgb}{0.93,0.00,0.00}
\definecolor{red3}{rgb}{0.80,0.00,0.00}
\definecolor{red4}{rgb}{0.55,0.00,0.00}
\definecolor{red}{rgb}{1.00,0.00,0.00}
\definecolor{rosybrown}{rgb}{0.74,0.56,0.56}
\definecolor{royalblue}{rgb}{0.25,0.41,0.88}
\definecolor{saddlebrown}{rgb}{0.55,0.27,0.07}
\definecolor{salmon1}{rgb}{1.00,0.55,0.41}
\definecolor{salmon2}{rgb}{0.93,0.51,0.38}
\definecolor{salmon3}{rgb}{0.80,0.44,0.33}
\definecolor{salmon4}{rgb}{0.55,0.30,0.22}
\definecolor{salmon}{rgb}{0.98,0.50,0.45}
\definecolor{sandybrown}{rgb}{0.96,0.64,0.38}
\definecolor{seagreen}{rgb}{0.18,0.55,0.34}
\definecolor{seashell1}{rgb}{1.00,0.96,0.93}
\definecolor{seashell2}{rgb}{0.93,0.90,0.87}
\definecolor{seashell3}{rgb}{0.80,0.77,0.75}
\definecolor{seashell4}{rgb}{0.55,0.53,0.51}
\definecolor{seashell}{rgb}{1.00,0.96,0.93}
\definecolor{sienna1}{rgb}{1.00,0.51,0.28}
\definecolor{sienna2}{rgb}{0.93,0.47,0.26}
\definecolor{sienna3}{rgb}{0.80,0.41,0.22}
\definecolor{sienna4}{rgb}{0.55,0.28,0.15}
\definecolor{sienna}{rgb}{0.63,0.32,0.18}
\definecolor{skyblue}{rgb}{0.53,0.81,0.92}
\definecolor{slateblue}{rgb}{0.42,0.35,0.80}
\definecolor{slategray}{rgb}{0.44,0.50,0.56}
\definecolor{slategrey}{rgb}{0.44,0.50,0.56}
\definecolor{snow1}{rgb}{1.00,0.98,0.98}
\definecolor{snow2}{rgb}{0.93,0.91,0.91}
\definecolor{snow3}{rgb}{0.80,0.79,0.79}
\definecolor{snow4}{rgb}{0.55,0.54,0.54}
\definecolor{snow}{rgb}{1.00,0.98,0.98}
\definecolor{springgreen}{rgb}{0.00,1.00,0.50}
\definecolor{steelblue}{rgb}{0.27,0.51,0.71}
\definecolor{tan1}{rgb}{1.00,0.65,0.31}
\definecolor{tan2}{rgb}{0.93,0.60,0.29}
\definecolor{tan3}{rgb}{0.80,0.52,0.25}
\definecolor{tan4}{rgb}{0.55,0.35,0.17}
\definecolor{tan}{rgb}{0.82,0.71,0.55}
\definecolor{thistle1}{rgb}{1.00,0.88,1.00}
\definecolor{thistle2}{rgb}{0.93,0.82,0.93}
\definecolor{thistle3}{rgb}{0.80,0.71,0.80}
\definecolor{thistle4}{rgb}{0.55,0.48,0.55}
\definecolor{thistle}{rgb}{0.85,0.75,0.85}
\definecolor{tomato1}{rgb}{1.00,0.39,0.28}
\definecolor{tomato2}{rgb}{0.93,0.36,0.26}
\definecolor{tomato3}{rgb}{0.80,0.31,0.22}
\definecolor{tomato4}{rgb}{0.55,0.21,0.15}
\definecolor{tomato}{rgb}{1.00,0.39,0.28}
\definecolor{turquoise1}{rgb}{0.00,0.96,1.00}
\definecolor{turquoise2}{rgb}{0.00,0.90,0.93}
\definecolor{turquoise3}{rgb}{0.00,0.77,0.80}
\definecolor{turquoise4}{rgb}{0.00,0.53,0.55}
\definecolor{turquoise}{rgb}{0.25,0.88,0.82}
\definecolor{violetred}{rgb}{0.82,0.13,0.56}
\definecolor{violet}{rgb}{0.93,0.51,0.93}
\definecolor{wheat1}{rgb}{1.00,0.91,0.73}
\definecolor{wheat2}{rgb}{0.93,0.85,0.68}
\definecolor{wheat3}{rgb}{0.80,0.73,0.59}
\definecolor{wheat4}{rgb}{0.55,0.49,0.40}
\definecolor{wheat}{rgb}{0.96,0.87,0.70}
\definecolor{whitesmoke}{rgb}{0.96,0.96,0.96}
\definecolor{white}{rgb}{1.00,1.00,1.00}
\definecolor{yellow1}{rgb}{1.00,1.00,0.00}
\definecolor{yellow2}{rgb}{0.93,0.93,0.00}
\definecolor{yellow3}{rgb}{0.80,0.80,0.00}
\definecolor{yellow4}{rgb}{0.55,0.55,0.00}
\definecolor{yellowgreen}{rgb}{0.60,0.80,0.20}
\definecolor{yellow}{rgb}{1.00,1.00,0.00}

\usepackage{graphicx}
\usepackage{subfigure}
\begin{document}

\title{ Polymer Adsorption on Curved Surfaces}
\author{Handan Arkin}
\email[E-mail: ]{Handan.Olgar@eng.ankara.edu.tr}
\affiliation{Institut f\"ur Theoretische Physik,
Universit\"at Leipzig, Postfach 100\,920, D-04009 Leipzig, Germany }
\affiliation{Department of Physics Engineering, Faculty of Engineering, Ankara University,
Tandogan, 06100 Ankara, Turkey}
\author{Wolfhard Janke$^{1}$}
\email[E-mail: ]{Wolfhard.Janke@itp.uni-leipzig.de}
\homepage[\\ Homepage: ]{http://www.physik.uni-leipzig.de/CQT.html}

\begin{abstract}
The conformational behavior of a coarse-grained  finite polymer chain near an attractive
spherical  surface was investigated by means of multicanonical Monte Carlo computer simulations.
In a detailed analysis of canonical equilibrium data over a wide range of sphere radius and temperature,
we have constructed entire phase diagrams both for non-grafted and end-grafted polymers. 
 For the identification of the conformational phases, we have calculated several energetic and  structural observables such as  gyration tensor based shape parameters and their fluctuations by canonical statistical analysis. 
 Despite the simplicity of our model, it qualitatively represents  in the considered parameter range  real systems that are studied in experiments. 
The work dicussed here could have experimental implications from 
protein-ligand interactions to designing nano smart materials.  

\end{abstract}

\maketitle

\section{Introduction}

The interaction of macromolecules with differently  shaped substrates is particularly important for interdisciplinary
research and nano-technological applications including, e.g., the fabrication of 
biosensors~\cite{Service} and peptide adhesion~\cite{Steiner} to metals~\cite{brown1, schulten1} or
semiconductors~\cite{whaley1, goede1, bachmann_sem}.
Gaining knowledge of structure formation for a variety of interfaces has therefore been a challenging subject of numerous experimental, theoretical 
and computational investigations. This includes thermodynamic studies of polymers at planar 
surfaces~\cite{eisenriegler_etal-1982,descas_etal-2004,mb-wj1,mb-wj2,mb-wj3,luettmer-strathmann_etal-2008,mb_wj-review-2008,moeddel1,ivanov_etal-2009,chen_etal-2009,wang_etal-2009,swetnam_allen2009,moeddel2,moeddel3,blavatska12,wuest_etal-2013,moeddel4,taylor_luettmer-strathmann-2014,austin17}, and
also under pulling force~\cite{Krawczyk, Bhattacharya}, and  at curved surfaces such as nano-tubes, nano-strings and nano-particles~\cite{Russel2003,Lundqvist,Gurevitch1,Gurevitch2,Vogel,Vogel2}.
Polymer adsorption on flat substrates plays an important role within a wide perspective. Due to the many possible applications, these ``hard-soft'' hybrid systems have been extensively 
studied from all aspects. 
For instance, employing a 
single-chain mean-field theory for polymers grafted to a flat surface has featured different morphologies for which, 
by controlling the self-assembly conditions, non-aggregated chains can coexist with micelles~\cite{micelles}. 
The understanding of the conformational properties of a polymer requires systematic studies because of the cooperative
effect of the monomers in response to different system conditions. The structuring  effect 
of an attractive substrate results in
a rich phase behavior caused by the competition between monomer-monomer and monomer-surface interaction.

By performing  Monte Carlo simulations for detailed  atomistic and  generic 
coarse-grained lattice and  
continuum models, many studies
have been done to investigate nano-particle--polymer interactions for different geometries, such as  cylinder 
and sphere~\cite{Netz}.
Using computer simulations, Barr and Panagiotopoulos~\cite{Barr} studied a system of polymers grafted to 
a spherical nano-particle in salt solution to gain insight into the conformational behavior of polymers
on curved surfaces.
Silver and gold nano-particles have also been considered experimentally
as catalyst for enhanced amyloid peptide fibrillation~\cite{abel1,abel2,abel3}.
Tanaka {\em et al.}~\cite{Tanaka1, Tanaka2} examined the freezing transition of compact polyampholytes, 
for both single and multiple chains.  
There has been a number of studies of these systems to determine the effects of surface charge 
densities~\cite{Ruckenstein} and solvent conditions on the morphologies of polymer chains. 
Furthermore adsorption of charged chains such as polyelectrolytes by oppositely charged
surfaces is also an important aspect in surface and 
colloidal science~\cite{Muthukumar1, Muthukumar2}. Because of the 
electrostatic attraction between chains and surfaces, a charged
chain tends to be adsorbed onto the surface. These studies are also extended to oppositely 
charged blocks on the chains~\cite{Messina}, and Dobrynin and Rubinstein~\cite{Rubinstein, Dobrynin} addressed typical 
adsorption regimes for a salt-free environment using scaling law 
arguments. The interaction between polyelectrolytes and small spheres of opposite charge is of interest for many problems such as interaction between polyelectrolytes and micelles or formation of nucleosomal complex between DNA and 
proteins~\cite{Netz1, Netz2}.  

Given the plethora of specific applications, it is important to complement such
detailed studies with investigations of generic models that focus on the most
characteristic parameters of the systems and can hence provide a broad overview
of the involved phenomena. In this spirit 
we have recently investigated the purely steric  confinement effect  of a  spherical cage on 
 a coarse-grained flexible polymer chain to determine the 
influence on the location of the collapse  and freezing transitions~\cite{Marenz}. 
Another hybrid system under consideration
was a polymer chain {\em inside\/} an attractive spherical cage for which we have 
constructed the  phase diagram
depending on the attraction strength of the sphere inner wall and 
 the temperature~\cite{Arkin2012,Arkin2013a} and investigated the ground-state properties~\cite{Arkin2012b}. 
We have also compared the results
 with the case of an attractive flat surface~\cite{Arkin2013b}.  
 Both  systems exhibit  a rich
phase behavior ranging from highly ordered, compact to extended, random coil structures. 

Here, we consider the opposite situation: A nano-sphere whose attractive {\em outer\/} spherical surface is the target for the
adsorbing polymer.  This problem  could
have practical implications for a broad variety of applications ranging from protein-ligand binding,
designing smart sensors to molecular pattern recognition~\cite{ bogner, balog, ike, gupta} and
for the discovery of new drugs that bind to specific receptors. 
Therefore it is interesting to study 
the adsorption of macromolecules
on different types of substrates and identify the conformational changes that a polymer can experience at the interface.
In this paper, we are going to investigate a simple coarse-grained polymer model interacting with a spherical
 surface of varying curvature by means of multicanonical Monte Carlo computer simulations.
 This method enables us to give an overview of  the different structural phases of a 
flexible polymer  chain over a wide range
of sphere radius and temperature. In a comparative study, we consider the two cases of  
 non-grafted and end-grafted polymer chains.  

The rest of the paper is organized as follows.  In Section II the model system is described in detail.
 Our model is a simple model that enables  changing parameters
  on a broad scale, which allows 
mapping to different   real systems that are considered in experiments.
The primary parameters that are scanned to obtain two-dimensional phase diagrams 
are the radius of the nano-particles and the temperature.
We kept the adsorption strength constant in this study (whereas we varied it
in another earlier study). 
 Then, in Section III the multicanonical Monte Carlo simulation method is briefly 
reviewed  and the measured observables are introduced, where special attention is paid to 
invariants of the gyration tensor.
  Section IV presents and discusses our main results, the phase diagrams   for
the  two systems  under consideration.  Finally, Section V concludes the paper with a summary of our findings.

\section{Model}
\subsection{Bead-stick polymer model}

The polymer chain is described by a generic, coarse-grained continuum   model for
  homopolymers which
has also been used for studies of heteropolymers in the frame of the hydrophobic-polar model~\cite{Stilinger1,Stilinger2}.
As in lattice models, the adjacent monomers are connected by rigid covalent bonds. Thus,
the distance is kept fixed and set to unity, fixing the length scale. The contact interaction
of lattice models
is replaced by a distance-dependent $12-6$ Lennard-Jones (LJ) potential,
\begin{equation}\label{eq:energy}
E_{\rm LJ}  =  4\epsilon_{\rm LJ}\sum_{i=1}^{N-2}\sum_{j=i+2}^{N}\left[\left(\frac{\sigma}{r_{ij}}\right)^{12}-\left(\frac{\sigma}{r_{ij}}\right)^{6}\right],
\end{equation}
accounting for
short-range excluded volume repulsion and long-range interaction of non-bonded monomers at
distance $r_{ij} = |\vec{r}_{i}-\vec{r}_{j}|$. Each summand in (\ref{eq:energy}) is minimized
for $r_{ij} =2^{1/6}\sigma$ where it contributes $-\epsilon_{\rm LJ}$ to $E_{\rm LJ}$. In the
simulations we set $\epsilon_{\rm LJ}$ to unity, fixing the energy scale, and choose $\sigma=1$.
This model was first employed in two dimensions~\cite{Stilinger1,Stilinger2} and later generalized to 
three-dimensional AB proteins~\cite{irb1,irb2,Arkin}, partially with modifications taking implicitly 
into account additional torsional energy contributions of each bond.
For consistency with our previous work \cite{Arkin,moeddel1,moeddel2,moeddel3,Arkin2012,Arkin2012b,Arkin2013a,Arkin2013b}
we kept a very weak bending energy 
$E_{\rm bend} = \varkappa \sum_{i=1}^{N-2}\left( 1-\cos\vartheta_{i}\right)$ 
with $\varkappa = 1/4$ and $\vartheta_{i}$ denoting the angle between adjacent bonds
($\cos \vartheta_{i} = 
\left( \vec{r}_{i+1}-\vec{r}_{i}\right)\cdot\left( \vec{r}_{i+2}-\vec{r}_{i+1}\right)$). 
For such a small bending stiffness, however, the statistical properties are hardly
distinguishable from a truly flexible ($\varkappa = 0$) polymer (see, e.g., Fig.~1 in Ref.\ \cite{marenz-knots}).

%


\subsection{Surface Interaction}

\begin{figure*}[t]
\centering
\hspace*{-12mm}
\includegraphics[width=9.0cm]{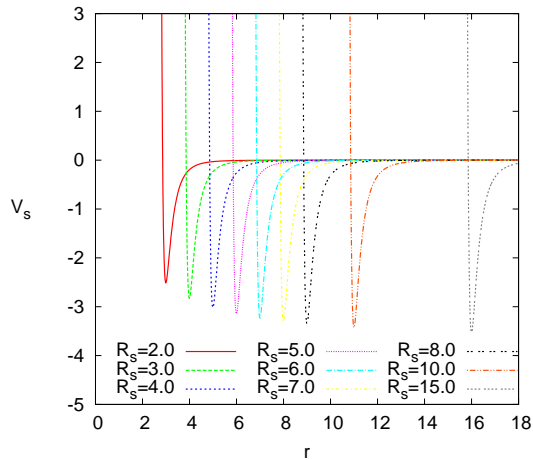}
\caption{\label{fig:pot} The functional dependence of the attractive sphere potential (\ref{eq5}) 
(with $\epsilon_{\rm s} = \sigma_{\rm s} = \epsilon = 1$) for different values of the sphere 
radius $R_{\rm s}$.  
}
\end{figure*}

In this work, we assume that the polymer chain interacts with an attractive spherical surface. 
As in our previous work \cite{Arkin2012,Arkin2012b,Arkin2013a,Arkin2013b}
the interaction of the polymer chain monomers and the attractive 
sphere is modeled by the surface energy $E_{\rm s} = \sum_{i=1}^N V_{\rm s}(r_i)$ where
%
%
%
%
%
\begin{widetext}
\begin{equation}\label{eq5}
	V_{\rm s}(r_i)=4 \pi \epsilon_{\rm s} \frac{R_{\rm s} }{r_i} \left\{  \frac{1}{5} \biggl [ \biggl (\frac{\sigma_{\rm s}}{r_i-R_{\rm s}} \biggr )^{10}-\biggl (\frac{\sigma_{\rm s}}{r_i+R_{\rm s}} \biggr )^{10}\biggr ]
	-\frac{\epsilon}{2}\biggl [ \biggl (\frac{\sigma_{\rm s}}{r_i-R_{\rm s}} \biggr )^{4}-\biggl (\frac{\sigma_{\rm s}}{r_i+R_{\rm s}} \biggr )^{4}\biggr ] \right\}.
\end{equation}
\end{widetext}
%
Here $R{\rm_s}$ is the radius of the sphere,
$r_i=(x_{i}^{2}+y_{i}^{2}+z_{i}^{2})^{1/2} \ge R{\rm_s}$ is the
distance of a monomer to the origin and $x_i, y_i, z_i$  are the coordinates of monomers, and
$\sigma_{\rm s}$,  $\epsilon_{\rm s}$, and $\epsilon$ are set to unity.  
The functional dependence of the potential $V_{\rm s}(r_i)$ is  
shown in Fig.~\ref{fig:pot} for selected $R{\rm_s}$ values which are used in the simulations.
For sufficiently large spheres and $r_i$ close to the surface, $r_i\approx R_{\rm s}$, we 
can neglect the terms $(\sigma/2R_{\rm s})^{10}$ and $(\sigma/2R_{\rm s})^{4}$ and
approximate \cite{Arkin2013b}
\begin{equation}
	V_{\rm s}(r_i)
	\approx 4\pi\epsilon_{s}\left[\frac{1}{5}\left(\frac{\sigma}{r_{i} - R_{\rm s}}\right)^{10}
	-\frac{\epsilon}{2}\left(\frac{\sigma}{r_{i} - R_{\rm s}}\right)^{4}\right],
\label{eq:V_s2}
\end{equation}
which is a standard $10-4$ Lennard-Jones potential with
$V_{\rm s}^{\rm min}=-4\pi\epsilon_{\rm s}(3/10)\epsilon^{5/3}R_{\rm s}/(R_{\rm s}+\sigma_{\rm s}\epsilon^{-1/6})$
at
$r_{i}^{\rm min}=R_{\rm s}+\sigma_{\rm s}\epsilon^{-1/6}$.

The total energy $E = E_{\rm LJ} + E_{\rm bend} + E_{\rm s}$ governs the statistical
properties at temperature $T$ respectively thermal energy $k_BT$, where $k_B$ is the 
Boltzmann constant. In the following we set $k_B$ to unity, fixing the temperature scale.
The most interesting phenomena result from the competition of intrinsic monomer-monomer and monomer-surface
wall interactions. For instance in the case of adsorption of polyelectrolyte chains onto 
oppositely charged interfaces, the electrostatic potential controls 
the competition of polymer-surface adsorption-desorption behavior. 

Our primary goal of this study is to obtain a broad overview of the phase diagram in the
$R_{\rm s}-T$ plane.
To make contact to specific experimental polymer-substrate systems  
one may identify the empirical (dimensionful) coupling parameters of, 
say, the Martini force field~\cite{MARTINI} with the (dimensionless) parameters of 
our coarse-grained model. For instance, from Table 1 in Ref.~\cite{Rodriguez}  we 
read off that methylene has $\epsilon_{\rm phys} = 104 \times 8.31/1000 = 0.86 
{\rm KJ/mol} $ which approximately corresponds to $\epsilon_{\rm LJ} = 1.0$ in our model.  
It follows that the 20mer (4 methylenes per bead)
considered in our study corresponds approximately to n-C80. Similarly, the substrate 
maps approximately onto a polystyrene colloidal sphere, but its adsorption propensity 
is weaker than that of carbon or silica. 


\section{Simulation Setup}
\subsection{Multicanonical method}
In order to obtain statistical results of sufficient accuracy we applied
the multicanonical (muca) Monte Carlo algorithm~\cite{MUCA1,MUCA2,MUCA3} (for reviews, see
Refs.~\cite{muca-review1,muca-review2,muca-review3}), where the energy distribution
is flattened artificially allowing, in principle, for a random walk of
successive states in energy space. This flattening is controllable and
therefore reproducible. To  this end, the Boltzmann probability is
multiplied by a weight factor $W(E)$, which in our case is a function
of the total energy $E = E_{\rm LJ} + E_{\rm bend} + E_{\rm s}$. Then the 
multicanonical probability for a state or conformation $\{ {\bf x} \}$
with energy $E(\{ {\bf x} \})$ reads $p_{\rm muca}(E)=\exp(-E/k_BT)W(E)$, 
up to an unimportant multiplicative factor. In order
to obtain a multicanonical or ``flat'' distribution, the initially
unknown weight function $W(E)$ has to be determined iteratively: In the
beginning, the weights $W^{(0)}(E)$ are set to unity for all energies
letting the first run be a usual Metropolis simulation which yields an
estimate $H^{(0)}(E)$ for the canonical distribution. This histogram is
used to determine the next guess for the weights, the simplest update
is to calculate $W^{(1)}(E)=W^{(0)}(E)/H^{(0)}(E)$. Then the next run
is performed with probabilities $p_{\rm muca}^{(1)}(E)=\exp(-E/k_BT)W^{(1)}(E)$
of states with energy $E$, yielding $H^{(1)}(E)$  and
$W^{(2)}(E)=W^{(1)}(E)/H^{(1)}(E)$, and so on. The iterative procedure
is continued until the weights are appropriate in a way that the
multicanonical histogram $H(E)$ is ``flat''. After having determined
accurate weights $W(E)$, they are kept fixed and following some
thermalization sweeps a long production run is performed, where
statistical quantities $O$ are obtained multicanonically,
$\langle O\rangle_{\rm muca}=\sum_{\{\bf x\}}p_{\rm muca}(E(\{{\bf x}\}))O(\{{\bf x}\})/Z_{\rm muca}$
with the multicanonical partition function
$Z_{\rm muca}=\sum_{\{\bf x\}}p_{\rm muca}(E(\{{\bf x}\}))$. The canonical statistics
is obtained by reweighting the multicanonical to the canonical
distribution, i.e., canonical expectation values are computed as
$\langle O\rangle=\langle OW^{-1} \rangle_{\rm muca}/\langle W^{-1} \rangle_{\rm muca}$.
For a recent review of these methodological aspects in the context of
polymer simulations, see Refs.~\cite{wj_wp2016, jz_mm_wj2016}.

\subsection{Observables}

To obtain as much information as possible
about the canonical equilibrium behavior, we define the following
suitable quantities $O$. Next to the canonical expectation
values $\left\langle O\right\rangle$, we also determine the fluctuations about these
averages, as represented by the temperature derivative $d\langle O \rangle/dT = \left( \left\langle O E \right\rangle -\left\langle O\right\rangle \left\langle E\right\rangle \right) /T^2$.
We use generic units, in which $k_B=1$.

 In order to identify conformational transitions, the specific heat (per monomer)
$C_V(T)=(\langle E^2\rangle-\langle E\rangle^2)/NT^2$ with
$\langle E^k\rangle=\sum_E g(E) E^k \exp(-E/T)/\sum_E g(E) \exp(-E/T)$ is calculated
from the density
of states $g(E)$. The density of states was found (up to an unimportant overall
normalization constant) by reweighting the multicanonical
energy distribution obtained with multicanonical sampling to the canonical distribution. Details are given in Ref.~\cite{Arkin}.

Apart from  the specific heat,
several structural quantities are of interest.
In order to check the structural  compactness of conformations or to identify the possible dispersion
of conformations because of adsorption, the radius of gyration  of the conformations is calculated.
The radius of gyration is a measure for the extension of the polymer
and defined by $R_{\rm g}^{2}\equiv\sum_{i=1}^{N} ( \vec{r}_i-\vec{r}_{\rm cm}) ^{2}/N=
\sum_{i=1}^{N}\sum_{j=1}^{N} ( \vec{r}_i-\vec{r}_{j}) ^{2} /2N^{2}$ with
$\vec{r}_{\rm cm}=\sum_{i=1}^{N}\vec{r}_i/N$ being the center-of-mass of the polymer.

We also calculated various  shape descriptors derived from the gyration tensor~\cite{Solc, Suter, Blavatska,Vymetal}  
which is defined as
\begin{equation}
S=\frac{1}{N} \left(   \begin{array}{ccc}
{\textstyle \sum_i (x_{i}-x_{\rm cm})^{2}} & \sum_i (x_{i}-x_{\rm cm})(y_{i}-y_{\rm cm}) & \sum_i (x_{i}-x_{\rm cm})(z_{i}-z_{\rm cm})\\
\sum_i (x_{i}-x_{\rm cm})(y_{i}-y_{\rm cm}) & \sum_i (y_{i}-y_{\rm cm})^{2} & \sum_i (y_{i}-y_{\rm cm})(z_{i}-z_{\rm cm})\\
\sum_i (x_{i}-x_{\rm cm})(z_{i}-z_{\rm cm}) & \sum_i (y_{i}-y_{\rm cm})(z_{i}-z_{\rm cm}) & \sum_i (z_{i}-z_{\rm cm})^{2}\end{array} \right).   
\end{equation}
Transformation to the  principal axis system diagonalizes $S$,  
\begin{equation}
S=  {\rm diag}(\lambda_{1}, \lambda_{2}, \lambda_{3}),
\end{equation}
where we assume  that the eigenvalues of $S$  are sorted in descending order,
i.e., $ \lambda_{1} \ge \lambda_{2} \ge \lambda_{3} $. 
The first invariant of $S$ gives the squared radius of gyration,

\begin{equation}
{\rm Tr} \,S= \lambda_{1} + \lambda_{2} + \lambda_{3} = R_{\rm g}^{2},
\label{eq:1st-invariant}
\end{equation}
which agrees with the definition given  above.
The second invariant  shape descriptor, or relative shape anisotropy,   is defined as
%
%
%
%
\begin{equation}
\kappa^{2}\equiv A_{3} = \frac{3}{2}\frac{{\rm Tr} \hat{S^{2}}}{({\rm Tr} S)^{2}}=1-3\frac{\lambda_{1}\lambda_{2}+\lambda_{2}\lambda_{3}+\lambda_{3}\lambda_{1}}{(\lambda_{1}+\lambda_{2}+\lambda_{3})^{2}},
\label{eq:2nd-invariant}
\end{equation}
where $\hat{S}=S-\frac{1}{3}({\rm Tr} S)I$ with unit tensor $I$.
It reflects both the symmetry and dimensionality of a polymer conformation. This parameter is limited between 
the values of 0 and
1. It reaches 1 for an ideal linear chain and drops to zero for highly symmetric conformations. For planar symmetric objects,
the relative shape anisotropy converges to the value 
of $1/4$~\cite{Solc, Suter, Blavatska, Vymetal, Blavatska2}. 


The
distance of the center-of-mass, $r_{\rm cm}$, of the polymer to the surface also provides clear evidence that the polymer
is freely moving or that it is very close to the surface and just adsorbed.
Another useful quantity is the mean number of monomers $\langle N_{\rm s} \rangle$
docked to the surface, which plays the role of an order parameter for the adsorption 
transition. A single-layer structure is formed if all
monomers are attached at the sphere; if none is attached, the
polymer is desorbed. The sphere potential
is a continuous potential, and in order to distinguish monomers
docked to  the sphere  from those not being docked, it is
reasonable to introduce a cutoff. We define a monomer $i$ as
being ``docked'' if $r_i-R_{\rm s}<r_c\equiv 1.2$. The corresponding measured
quantity is the average number  $\left\langle N_{\rm s}\right\rangle$ of monomers docked to the
surface. This can be expressed as
$N_{\rm s}=\sum_{i=1}^N \Theta( r_c-r_i) $,
where $\Theta(r)$ is the Heaviside step function.

\begin{figure*}[t]
\vspace*{-24mm}
\subfigure[]{\includegraphics[width=11.5cm]{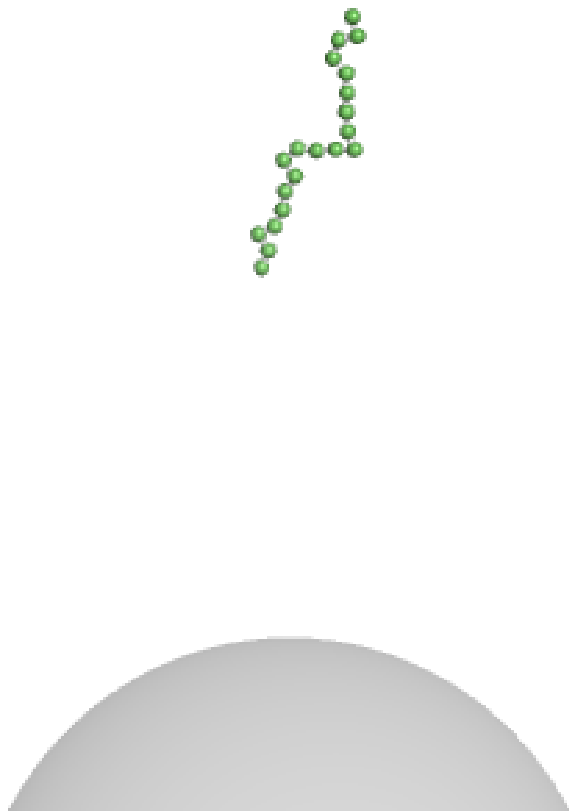}}
\subfigure[]{\includegraphics[width=6.0cm]{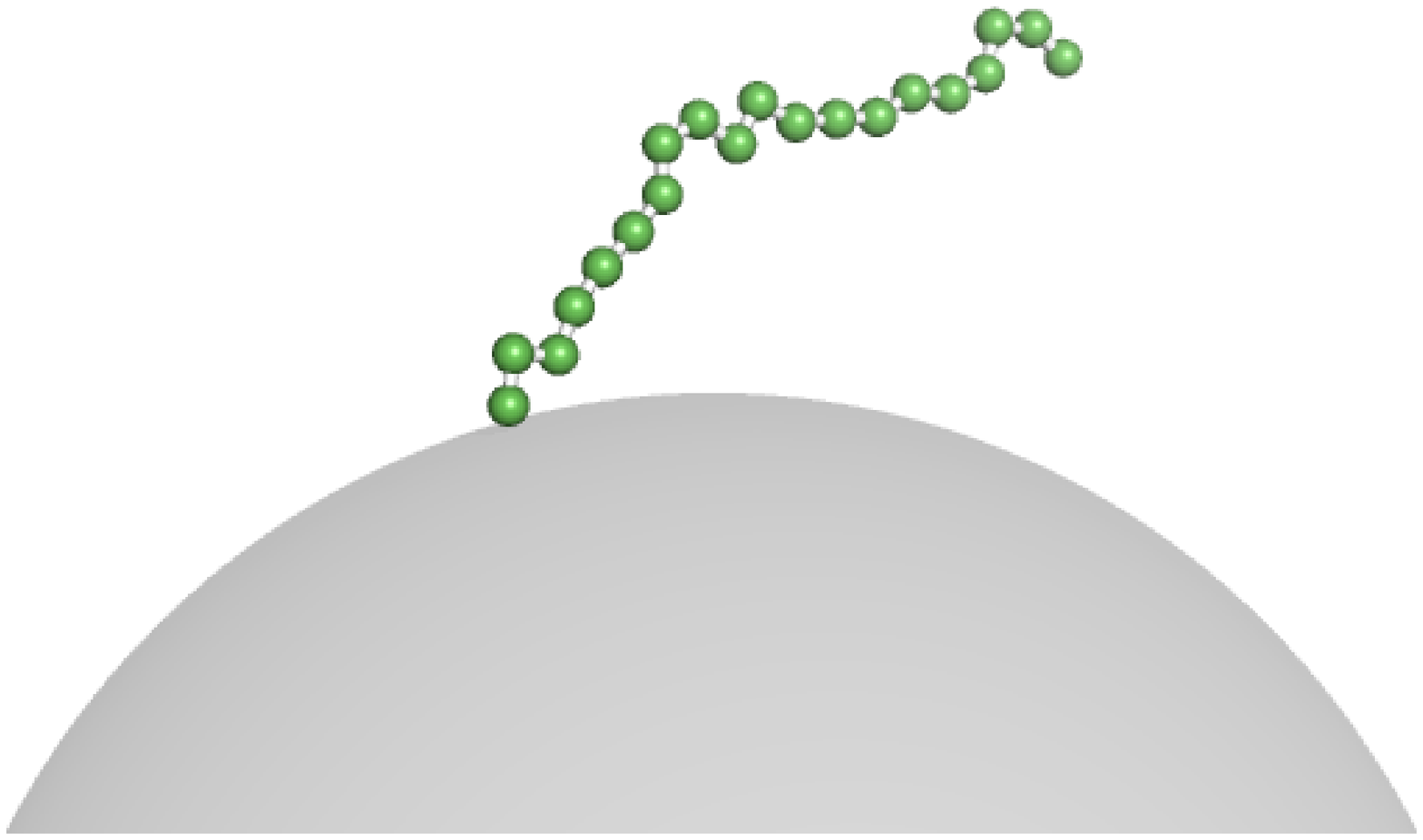}}
\caption{\label{fig:start} (Color online)  Start configurations of the simulations: (a) non-grafted and 
(b) end-grafted polymers of length $N=20$.
}
\end{figure*}

\subsection{Computational details}
In our simulations, the polymer chain length is $N=20$ and
we set $\epsilon=1.0$ in the surface potential (\ref{eq5}) large enough 
to allow adsorption of  the polymer  to  the sphere surface.  
We consider two different situations, one is the case where 
the polymer is allowed to move freely in the space around the sphere over a distance $L=60 - R{\rm_s}$ 
from its surface (i.e., the nano-sphere of radius $R{\rm_s}$ is 
centered in a spherical container of radius 60 with a purely steric wall), 
which is called the ``free'' or
``non-grafted'' case, and in the second case it is grafted with one end to the
surface (``end-grafted'').

 We have done  simulations with different sizes of the 
sphere.   
The random  initial configurations for the non-grafted and end-grafted cases  of the simulation are  sketched in Figs.~\ref{fig:start}(a) and (b).
The  total energy   of the system is  composed of the   pure  polymer chain energy $E_{\rm LJ} + E_{\rm bend}$ and the polymer chain attractive sphere  
interaction energy $E_{\rm s}$.
The initial configuration of the polymer chain is randomly generated.
For the determination of the multicanonical weights we performed $200$ iterations with
at least $10^5$ sweeps each.
In the production period, $10^8$
sweeps were generated to have reasonable statistics for estimating the thermodynamic quantities.
Statistical errors are estimated with the standard Jackknife technique~\cite{jack1,jack2,wj_lviv}.


\section{Results }

\begin{figure*}[h]
	\subfigure[]{\includegraphics[width=8.5cm]{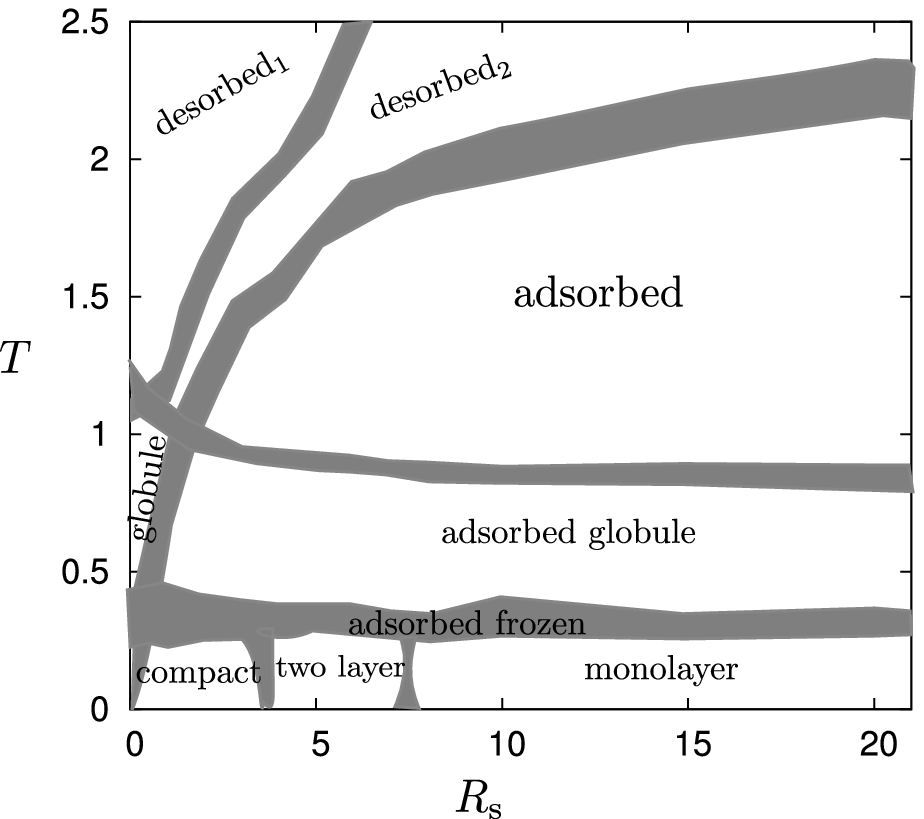}}\hspace*{10mm}
\subfigure[]{\includegraphics[width=8.3cm]{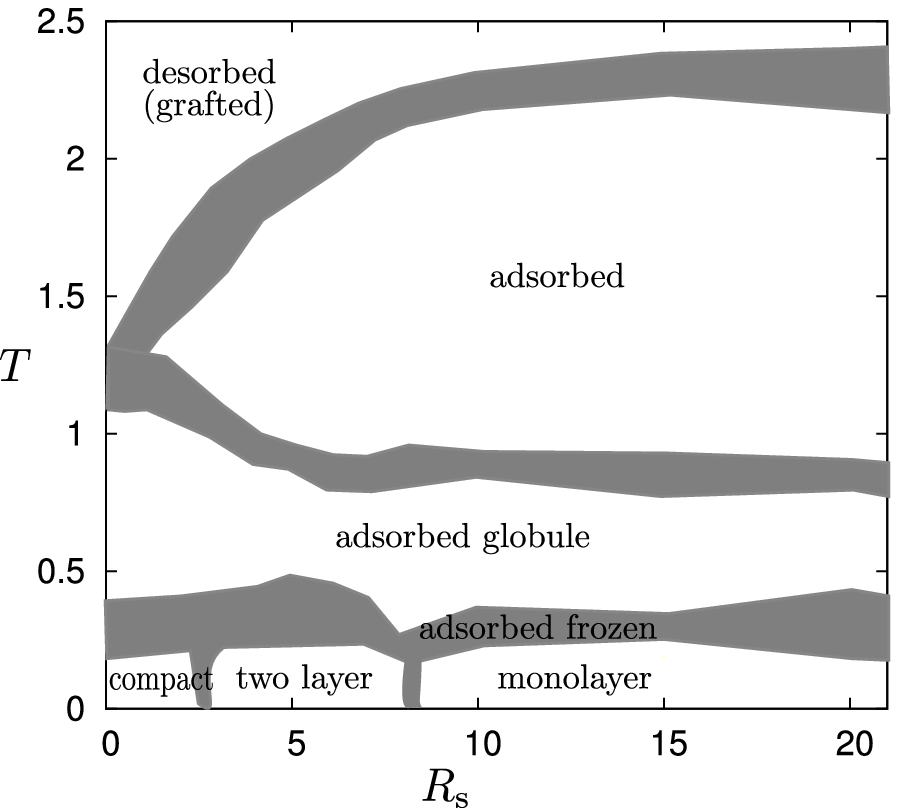} }
\caption{\label{fig:phase} The phase diagram of the homopolymer-attractive spherical surface  system 
 for (a) a non-grafted  and
 (b) an end-grafted polymer
as obtained from extensive multicanonical simulations. The grey bands separate the
individual conformational phases. The band width shows the variation
of the peaks of temperature derivatives of different energetic and structural observables
which have been analyzed simultaneously.  In our simulations, the polymer chain 
length is $N=20$ and we set $\epsilon=1.0$ for the surface attraction strength. }
\end{figure*}

\subsection{Phase diagrams}  

To give an overview at the beginning, we start by presenting in Fig.~\ref{fig:phase} the main result
of our study: The phase structure for (a) non-grafted and (b) end-grafted polymers 
derived from all of our observables as summarized by phase diagrams in the 
$R{\rm_s} - T$ plane. 
These phase diagrams are constructed by combining all the informations coming from the 
canonical expectation values of our observables
and their temperature derivatives described in more detail in the next two subsections. 
Some of our observables exhibit a peak at all of the transitions in the phase diagram, 
while others are only sensitive to one of them. For example, the collapse transition 
line at $T \approx 0.8$ is seen quite clearly from the peak in the temperature derivative
of the canonical expectation value of  the radius of gyration (\ref{eq:1st-invariant})
and as a small shoulder in the invariant shape anisotropy (\ref{eq:2nd-invariant}). 
Naturally, this is further 
complemented by information coming directly from the eigenvalues of the gyration tensor. 
On the other hand, the adsorption line running roughly between $T \approx 1.5$ (small $R_{\rm s}$) 
and $T \approx 2.5$ (large $R_{\rm s}$) is most clearly constructed by looking at 
the mean distance of the center-of-mass of the polymer to the surface and
the mean number of monomers adsorbed onto the surface, which plays the role of an order 
parameter for this transition.  
Since we are dealing with a finite system, it is not possible to determine the transition lines precisely: 
The transition lines still vary with chain length $N$ and the observables have broad peaks. Thus 
we have a certain band width which approximately covers the different peaks in the observables 
(and strictly speaking one should talk of ``pseudo-transitions'' instead of ``transitions'' 
and ``pseudo-phases'' instead of ``phases'', but for brevity we will suppress the attribute 
``pseudo'' in the following). 

In the phase diagrams the radius of the sphere increases from left to right and 
the temperature increases from bottom to top. The grey
bands separate the individual conformational phases.  
For high temperature, the polymer behaves in both cases similarly to a free polymer
where the typical conformations are desorbed and extended random coils. 
In the non-grafted case with small sphere radius $R{\rm_s}$,
decreasing the temperature causes the (three-dimensional) collapse transition into 
globular conformations which are still in the desorbed phase. But below the freezing 
transition all the compact conformations 
are adsorbed. In contrast, for the end-grafted case
all conformations are already adsorbed below the collapse transition. 
There is no desorbed globule phase in the
 grafted phase diagram.  
One more difference  occurred also in the high-temperature desorbed phase. In the non-grafted case some structural 
observables give  indication for some changes in the desorbed phase. When we carefully analyze the conformations we see that 
 those in the ``desorbed$_1$'' phase are far away from the sphere surface while the conformations in the ``desorbed$_2$'' 
 phase are almost adsorbed. Thus they feel 
very strongly the surface effect. Because of the grafting, this is not the case for an end-grafted polymer.    
Increasing the sphere radius approximately to $R_{\rm s} \approx 7.0$ leads to a very fast increase in the adsorption 
transition  temperature, 
but after this value it increases slowly. The adsorption transition separates the regions of desorbed and adsorbed phases.    
Besides the collapse, adsorption, and freezing transitions, the most pronounced transition is the layering transition which occurs for low temperatures at 
$R{\rm_s} \approx 7.0$ and separates the region of planar conformations which are monolayers of totally 
adsorbed conformations from the two-layer conformations. 
Another low-temperature transition is coming into play at $R_{\rm s} \approx 3.0$,
where two-layer conformations change to compact conformations (which look almost like a sphere).

\begin{figure*}[t]
\centering
\subfigure[]{\includegraphics[width=4.0cm]{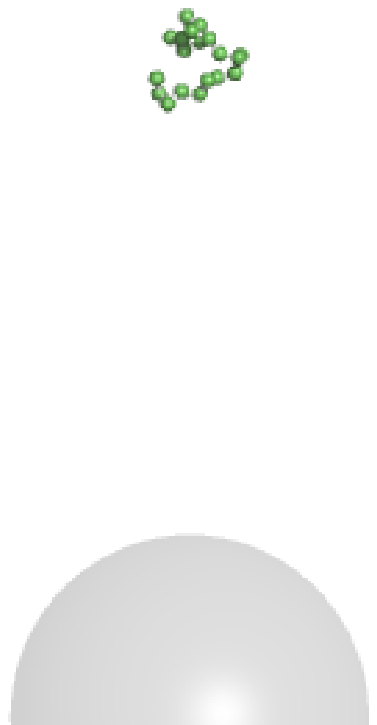}}
\subfigure[]{\includegraphics[width=4.0cm]{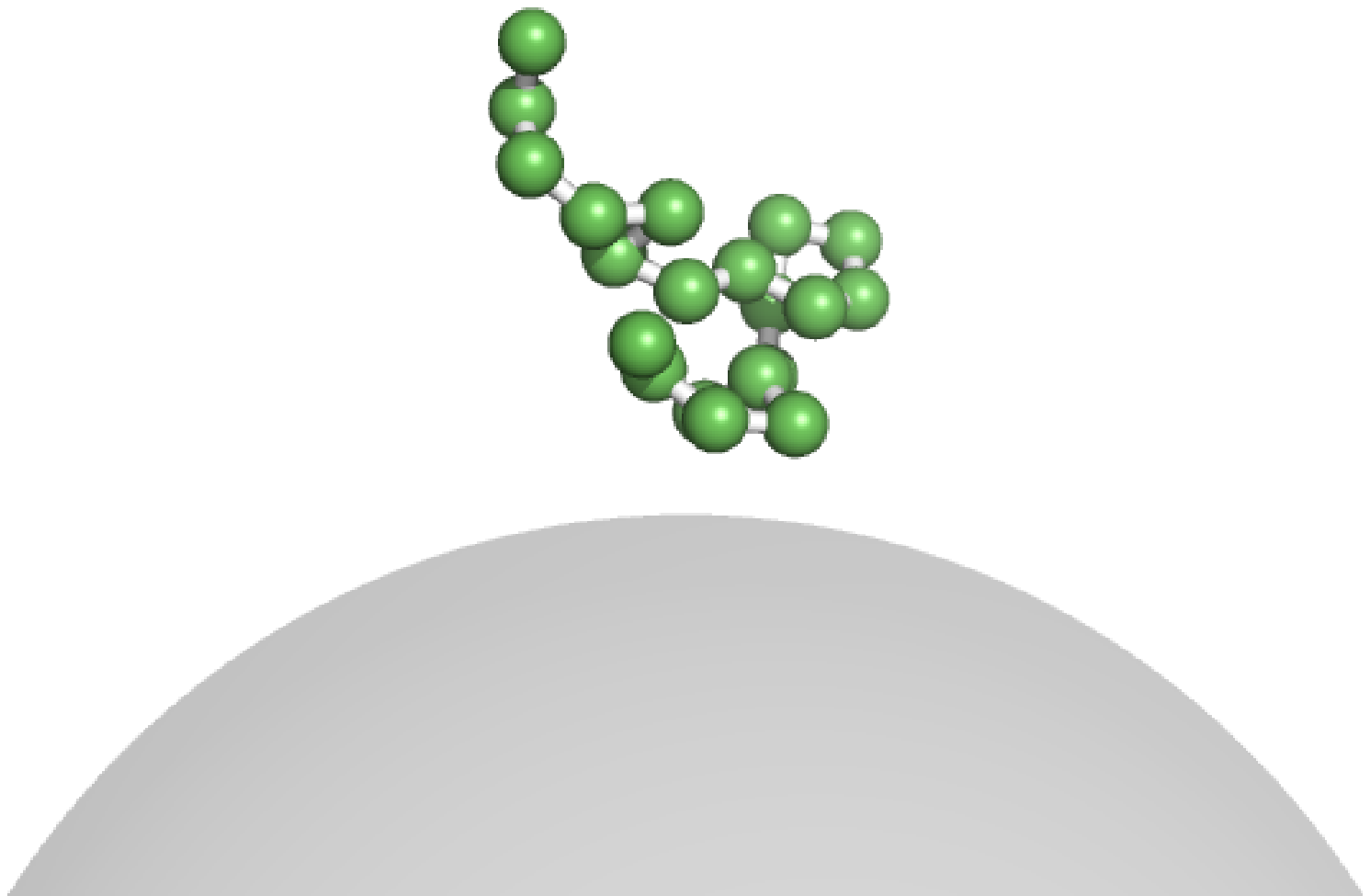}}
\subfigure[]{\includegraphics[width=4.0cm]{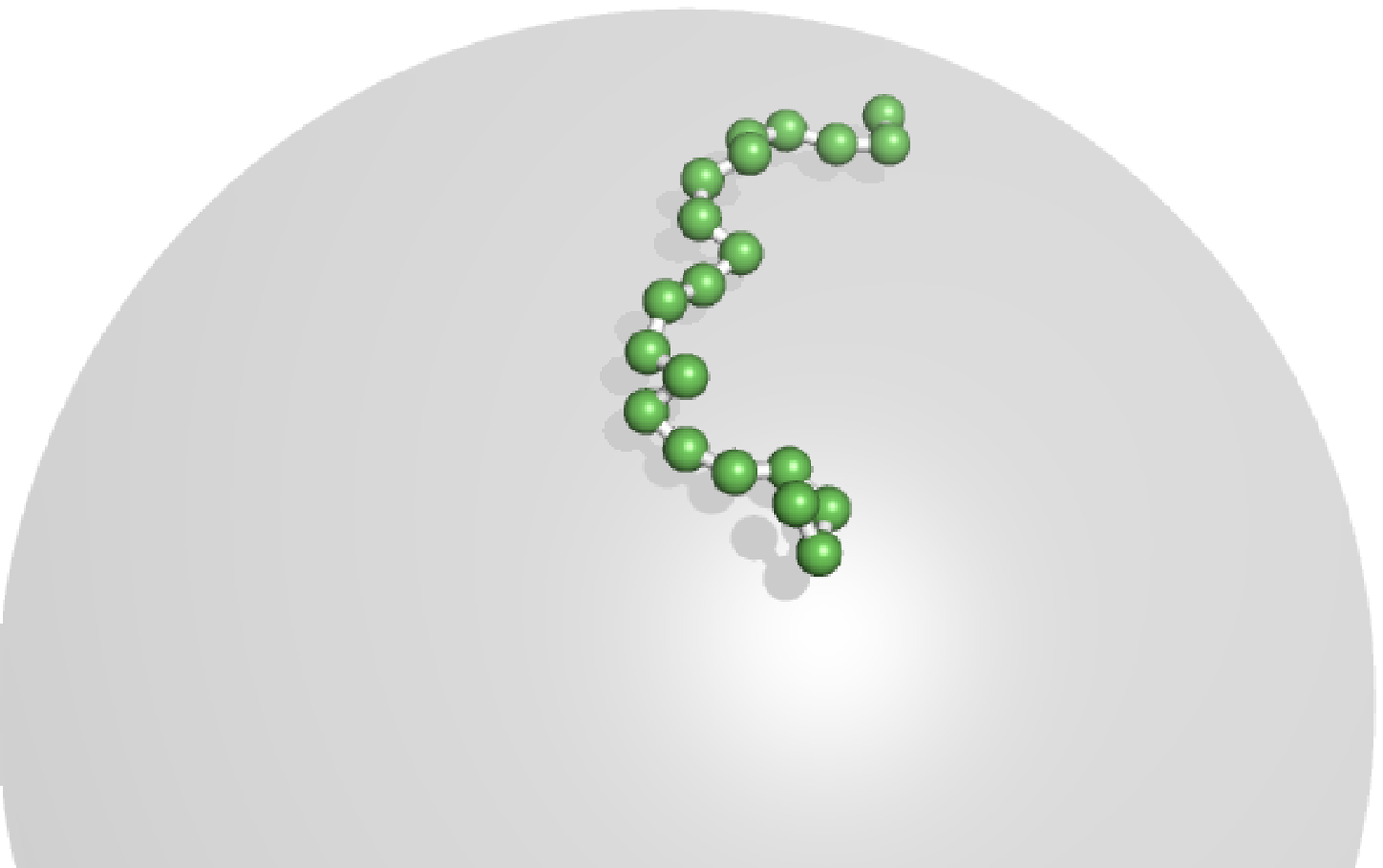}}
\\
\subfigure[]{\includegraphics[width=4.0cm]{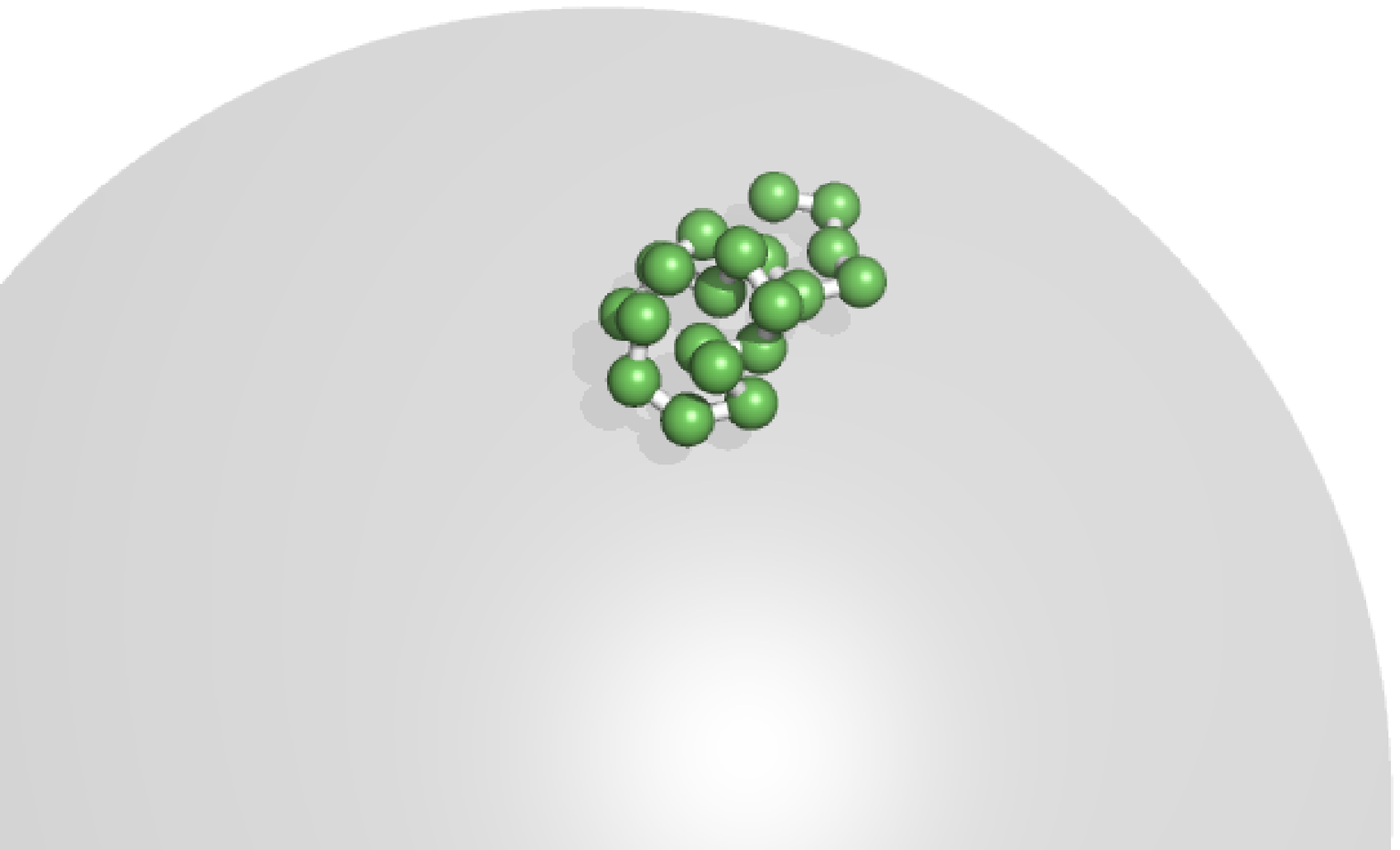}}
\subfigure[]{\includegraphics[width=4.0cm]{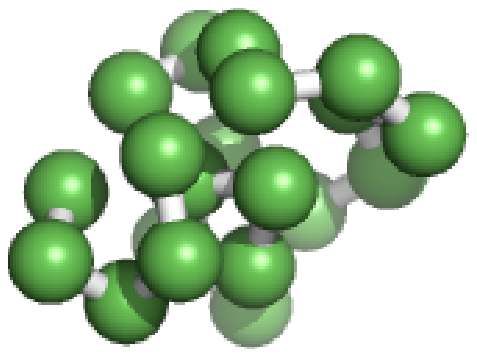}}
\subfigure[]{\includegraphics[width=4.0cm]{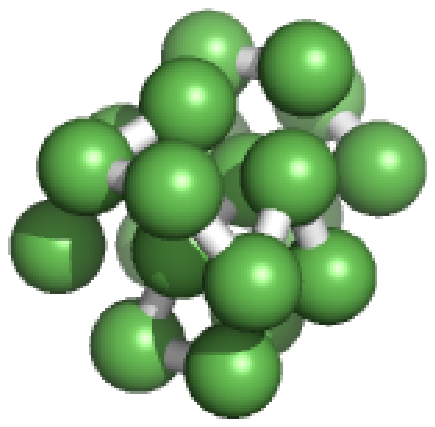}}
\\
\subfigure[]{\includegraphics[width=4.0cm]{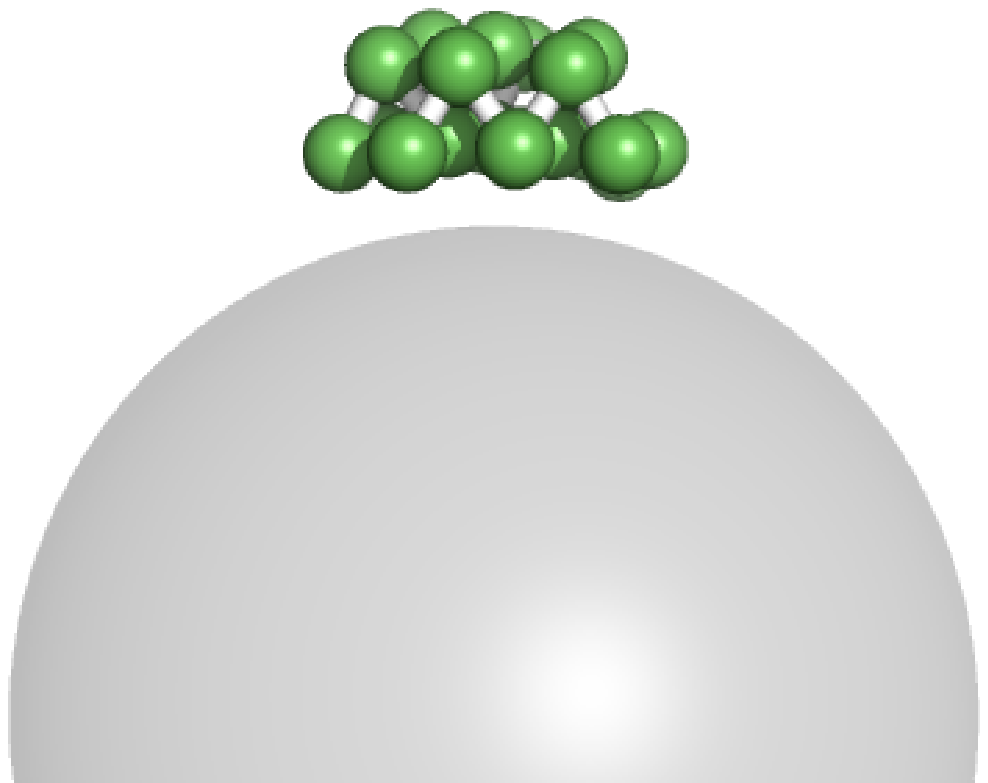}}
\subfigure[]{\includegraphics[width=4.0cm]{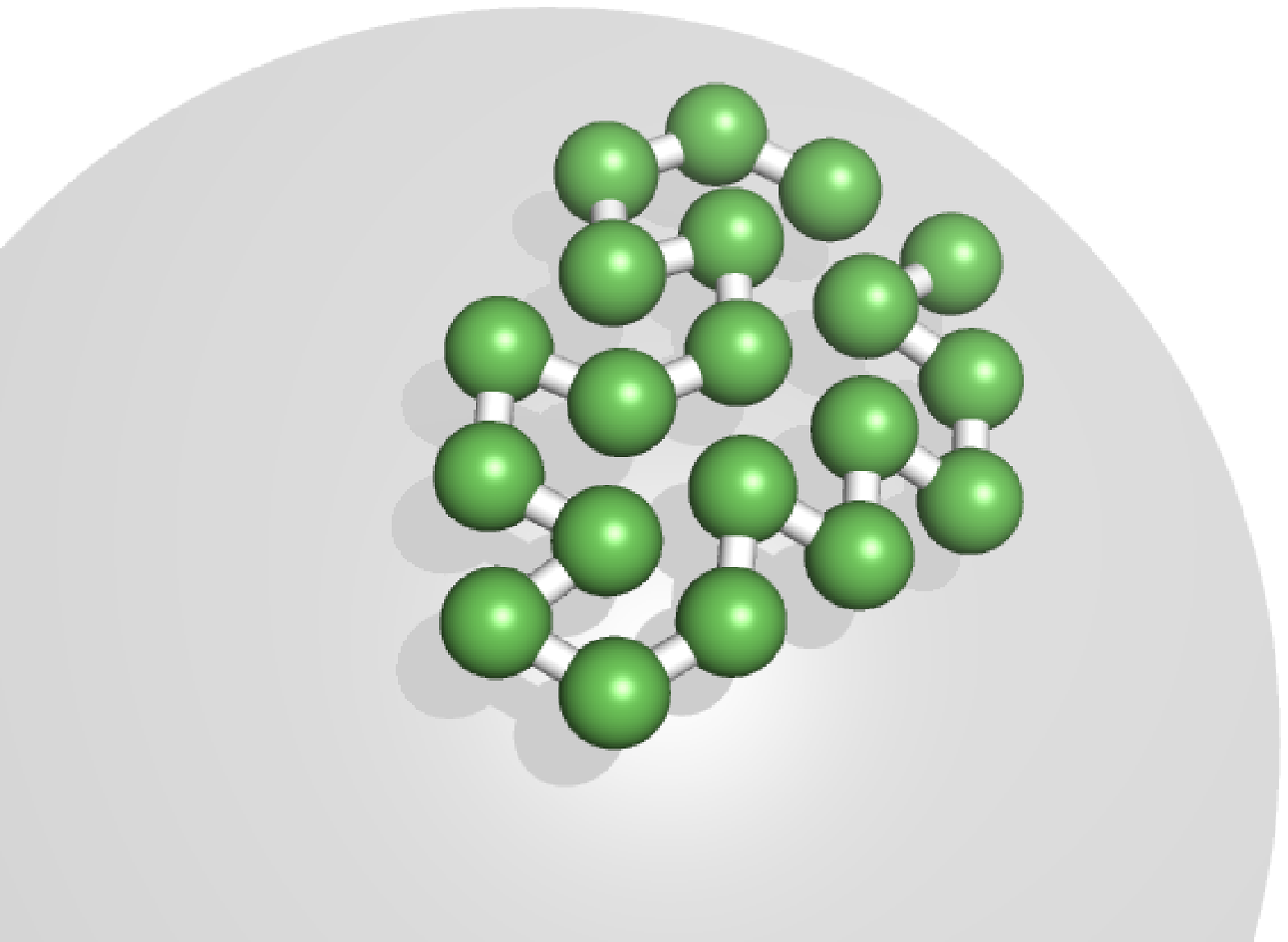}}
\caption{\label{fig:conf} (Color online) Typical   conformations for the regions (a) desorbed$_1$,  (b) desorbed$_2$,
(c) adsorbed, (d) adsorbed globule, (e) globule, (f) compact, (g) two layer, and (h) monolayer in the phase diagram for a non-grafted polymer. 
}
\end{figure*}

The representative conformations that predominate in the different structural phases 
are depicted in Fig.~\ref{fig:conf} for the case of a non-grafted polymer.
The observed structural phases for this case can be briefly summarized as follows:

\begin{description}

\item[Desorbed$_1$:]  Random coil structures with no surface contacts. These
conformations freely circulate in the simulation space and are far away from
the surface of the sphere [Fig.~\ref{fig:conf}(a)].

\item[Desorbed$_2$:]  Desorbed conformations, but they are  almost adsorbed.
The conformations feel the influence of the surface [Fig.~\ref{fig:conf}(b)].

\item[Adsorbed:] Partially adsorbed, extended conformations [Fig.~\ref{fig:conf}(c)].

\item[Adsorbed Globule:] Partially adsorbed  conformations  [Fig.~\ref{fig:conf}(d)].

\item[Globule:] Desorbed globule  conformations.
These conformations are only seen in the non-grafted phase diagram [Fig.~\ref{fig:conf}(e)].
\item[Compact:] Partially adsorbed, globular conformations like a drop on the wall
of the sphere [Fig.~\ref{fig:conf}(f)].

\item[Two Layer:] Partially adsorbed, compact conformations. These are two-layer
structures. The lower layer of the conformations is adsorbed  and lies on the wall
of the sphere [Fig.~\ref{fig:conf}(g)].

\item[Monolayer:] Completely adsorbed, compact conformations. These single-layer
structures lie on the surface of the sphere and fit the sphere wall
perfectly [Fig.~\ref{fig:conf}(h)].

\end{description}

In the following two sections we will discuss in more detail
how these phase diagrams have been obtained by analyzing energetic and
structural observables.

\subsection{Energetic fluctuations}  

\begin{figure*}
\vspace*{9mm}
\hspace*{-8mm} 
\subfigure[]{\includegraphics[width=8.9cm]{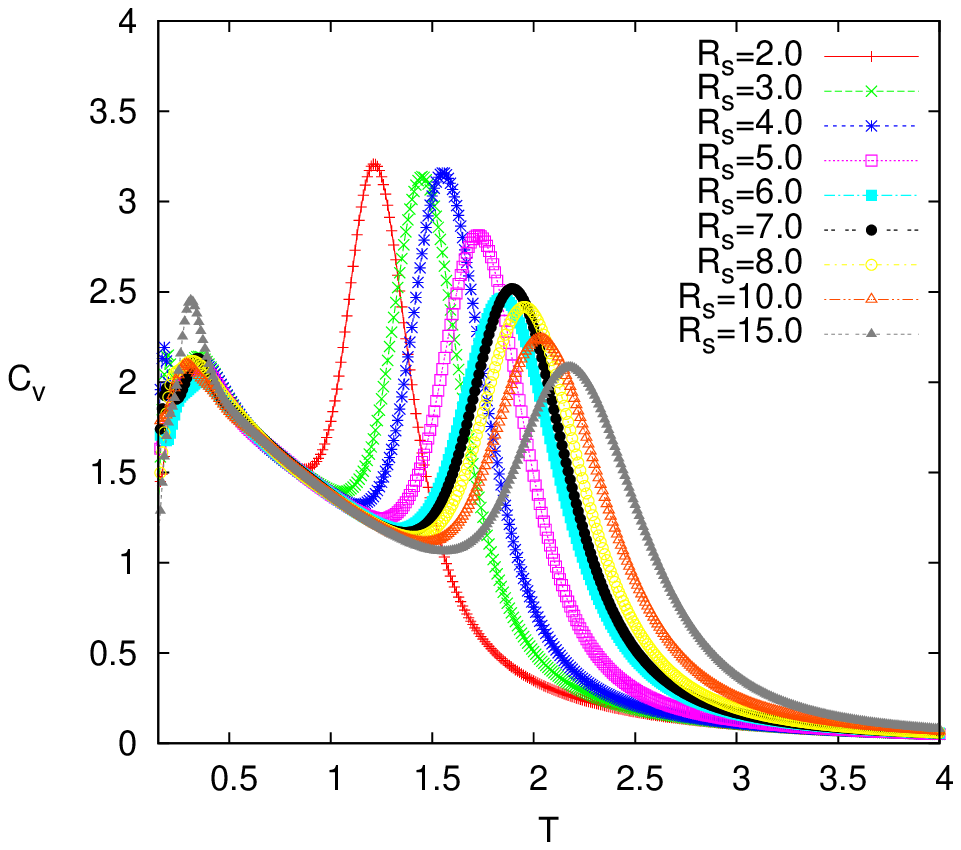}}
\subfigure[]{\includegraphics[width=8.9cm]{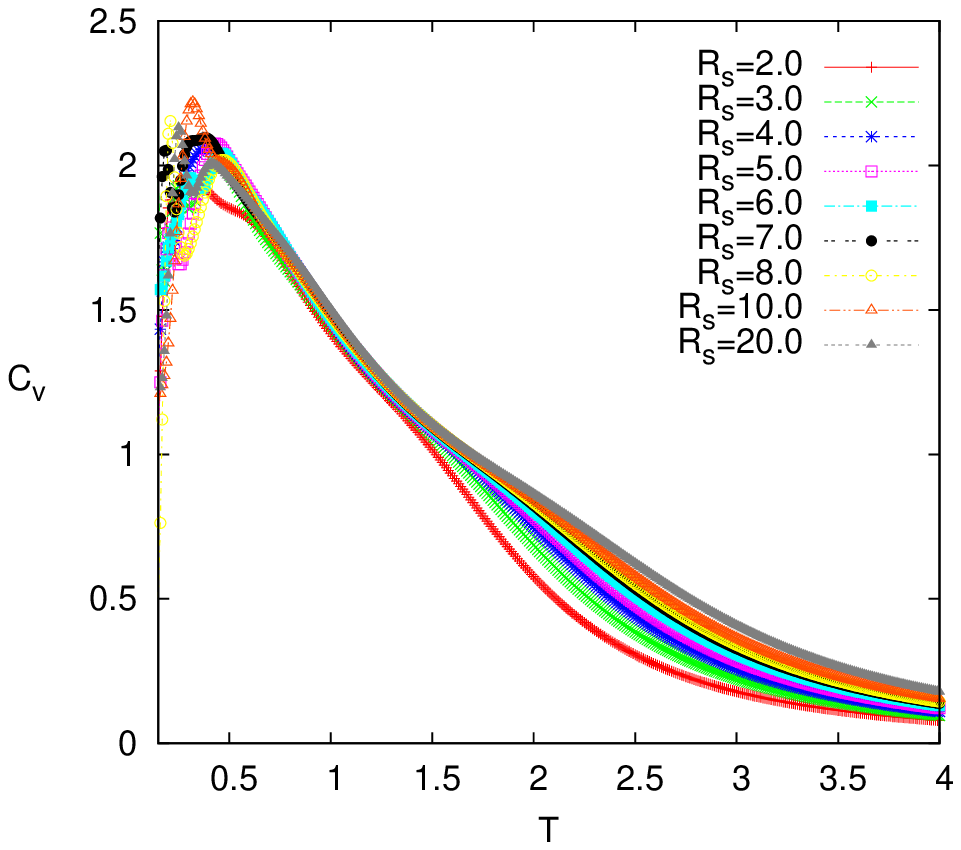}}
\caption{\label{fig:cv} (Color online) Specific heat as a function of temperature for different 
sphere radii $R{\rm_s}$ for the (a) non-grafted and (b) end-grafted case (polymer length $N=20$, 
surface attraction strength $\epsilon=1.0$). 
}
\end{figure*}

Figure~\ref{fig:cv} displays the specific-heat curves $C_V(T)$ as a function of temperature $T$
for different values of sphere radius $R{\rm_s}$ for (a) the non-grafted and (b) the end-grafted case.
In both cases the specific heat signals two transitions:
one is the low-temperature transition which is
almost at the same temperature  \textbf{ ($T \approx 0.3$)} for all different $R{\rm_s}$ values.
This is the freezing transition which does not differ much for non-grafted and end-grafted chains. 
The second transition is quite pronounced in the non-grafted case but
exhibits only a weak signal (a shoulder at $T \approx 2.0$) 
for the grafted polymer. 
This is the adsorption transition, which separates desorbed and
adsorbed conformations. It comes into play at higher temperatures than the 
freezing transition and depends quite strongly on the sphere radius.
This is consistent with previous observations that
for non-grafted polymers of finite length this transition has a first-order-like signature
(which eventually crosses over to second-order-like in the infinite chain-length 
limit) \cite{eisenriegler,Netz2,moeddel3},
whereas for grafted polymers it always looks like a continuous transition. 
 In both cases, increasing the sphere radius 
causes an increase in 
the adsorption transition temperature.

\subsection{Structural parameters and fluctuations}  
\subsubsection{Radius of gyration $R_{\rm g}$}

\begin{figure*}[h]
\hspace*{-8mm} 
\subfigure[]{\includegraphics[width=8.9cm]{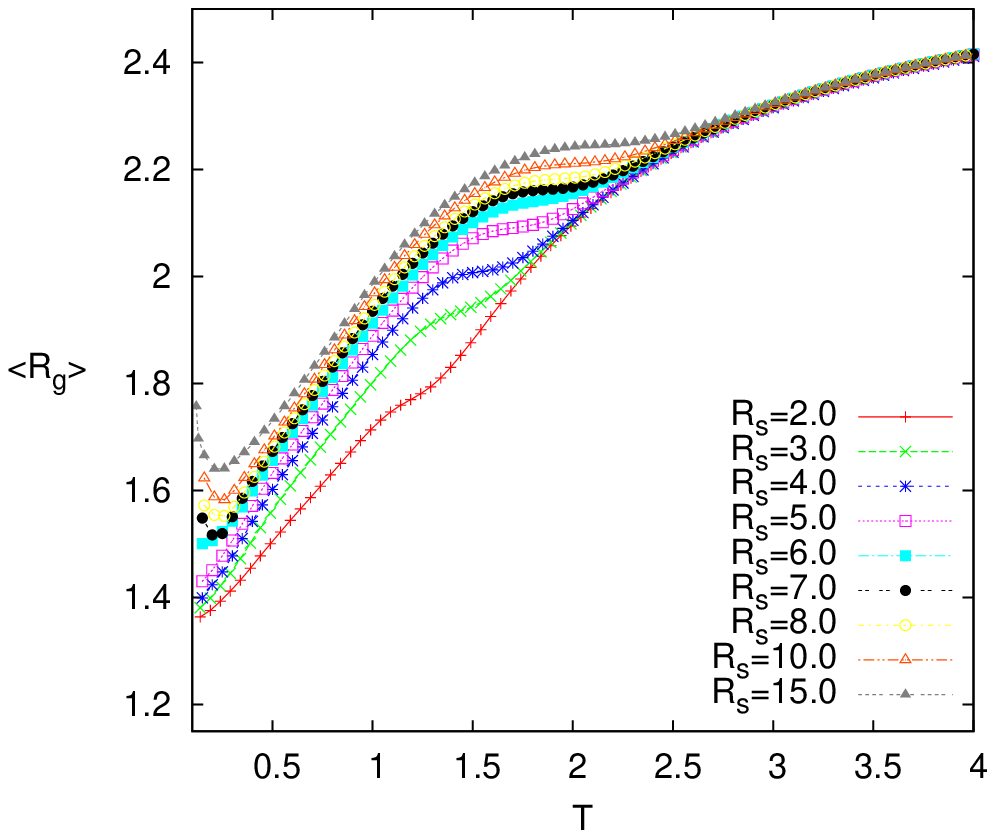}}
\subfigure[]{\includegraphics[width=8.9cm]{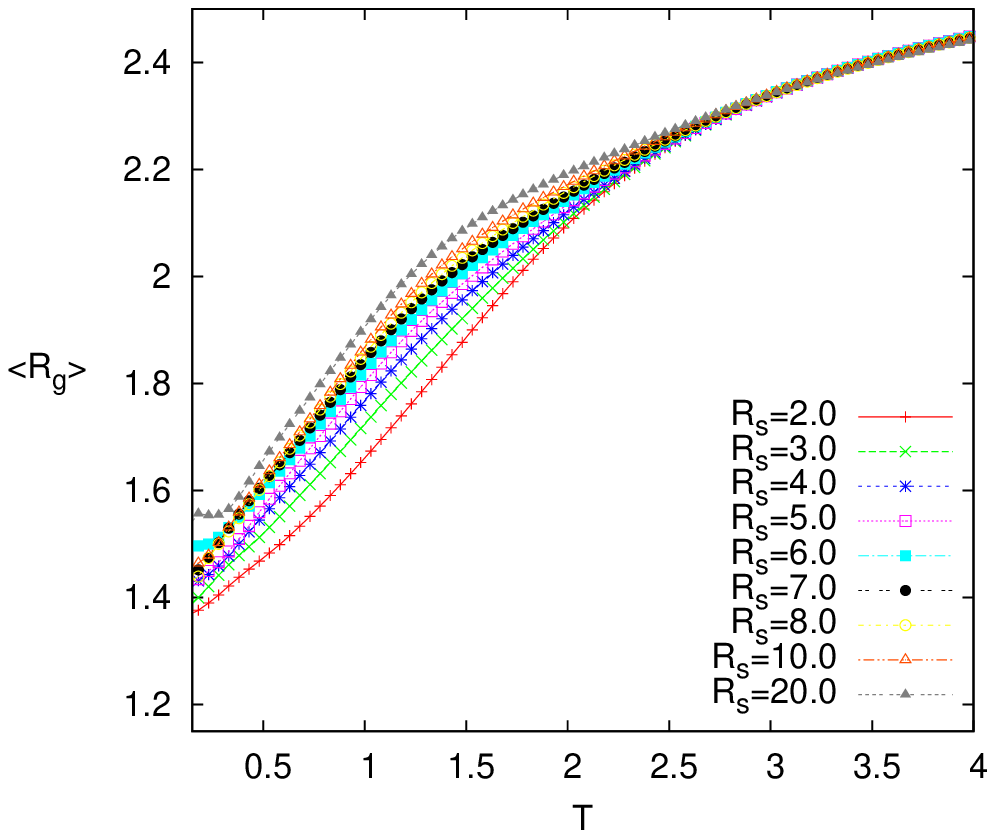}}
\hspace*{-8mm} 
\subfigure[]{\includegraphics[width=8.9cm]{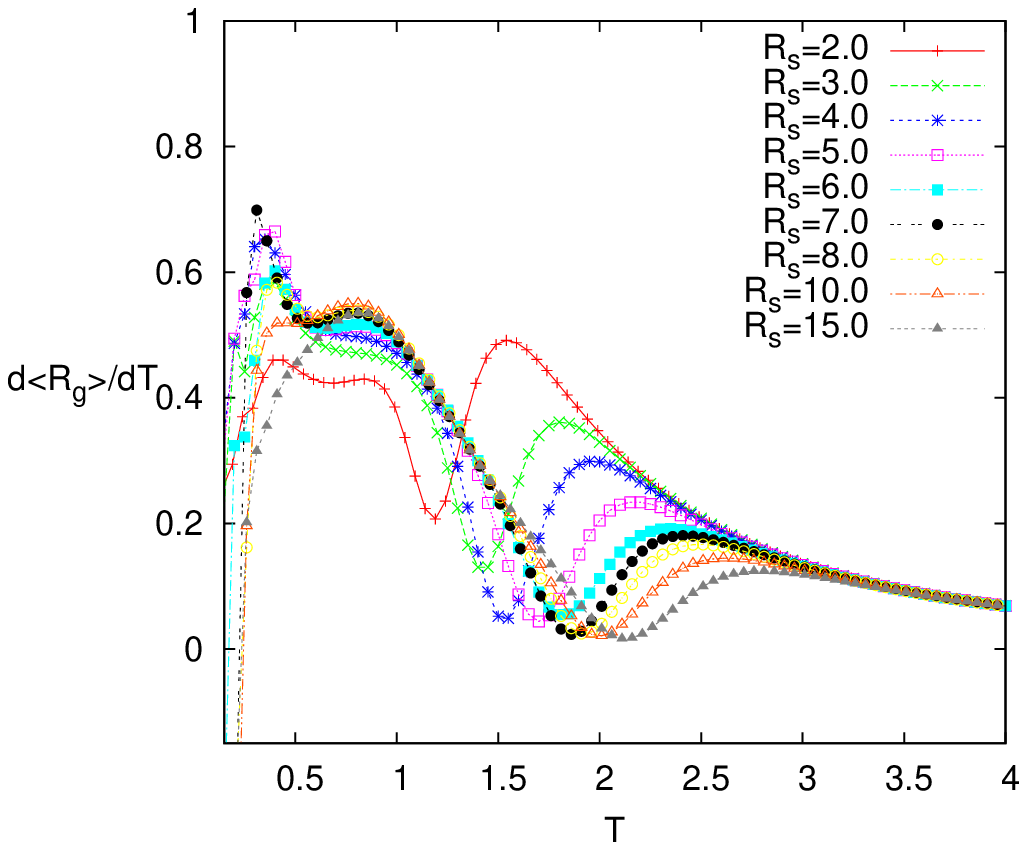} }
\subfigure[]{\includegraphics[width=8.9cm]{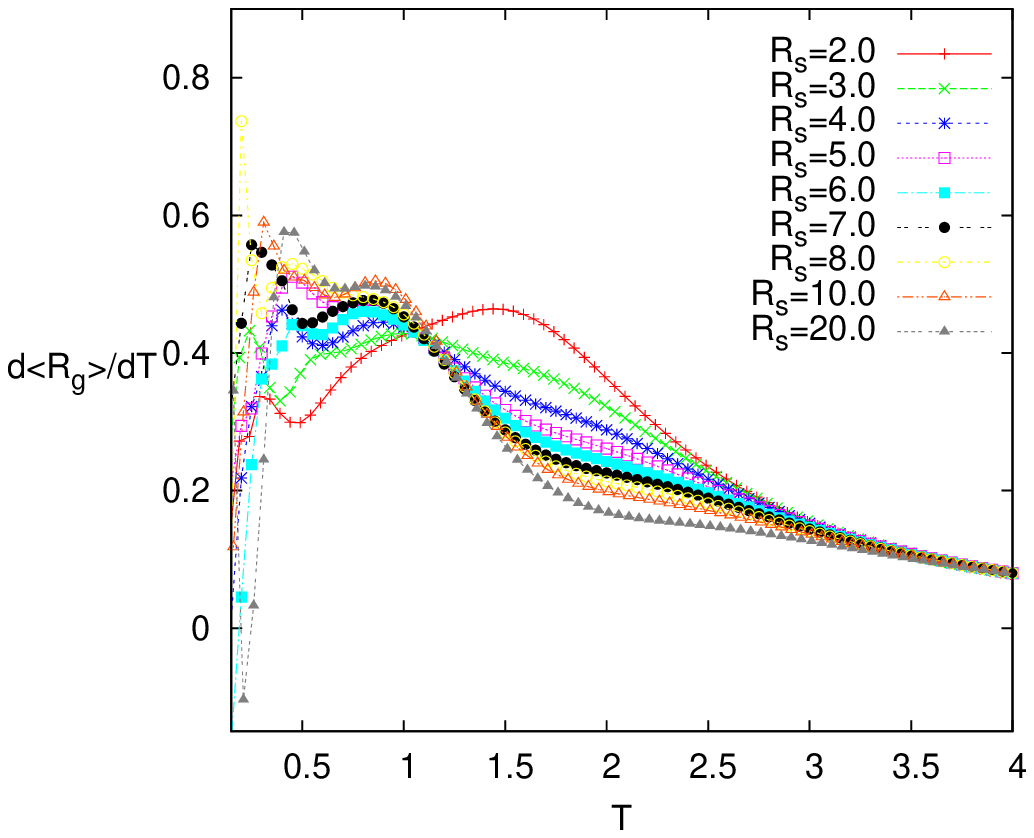}}
\caption{\label{fig:rgy} (Color online) The canonical expectation value of the radius of 
gyration $\langle R_{\rm g} \rangle$ for the (a) non-grafted and (b) end-grafted case, 
and (c), (d) the corresponding temperature derivatives, for different sphere 
radii $R{\rm_s}$ (polymer length $N=20$, surface attraction strength $\epsilon=1.0$).
}
\end{figure*}

The radius of gyration $\langle R_{\rm g}\rangle$ (the first invariant of 
the gyration tensor) and its temperature derivative  $d\langle R_{\rm g}\rangle/dT$ 
are shown in Fig.~\ref{fig:rgy} as a function of temperature for both the non-grafted and 
end-grafted cases, respectively. 
For small values of the sphere radius $R_{\rm s} = 0.5, 1.0$, the most compact conformations 
occur in the low-temperature region with an average $\langle R_{\rm g}\rangle \approx 1.23  $ 
(data not shown). Slightly increasing the $R_{\rm s}$ value causes 
an increase in the average 
$\langle R_{\rm g} \rangle $ value to about  $1.4$.  
Increasing  the $R_{\rm s}$ parameter further, the curve for  $R_{\rm s}=7.0$ of 
 Fig.~\ref{fig:rgy}(a) in  the non-grafted case
 has a minimum behavior at low temperatures. 
 As a function of temperature the radius of gyration is
 monotonically increasing for all
$R_{\rm s}$ values except beyond $R_{\rm s}=7.0 $,
where the layering transition occurs.
Supporting information is also gained from the relative shape anisotropy parameter 
in Fig.~\ref{fig:rgK}.  
If we now look at the temperature derivative of the radius of gyration in Fig.~\ref{fig:rgy}(c),
we detect three maxima for each $R_{\rm s}$ curve. 
The first peak 
at low temperatures ($T \approx 0.3$) indicates the freezing
transition quite clearly, the second peak around $T \approx 0.8$ can be identified with the
(two-dimensional) collapse transition, and the third, strongly moving peak in the
region $T \approx 1.5-3.0$ signals the adsorption transition. For the end-grafted case
these signals are generally weaker. In  Fig.~\ref{fig:rgy}(d), the first two maxima are 
still discernable, but the adsorption transition is hardly reflected.

\subsubsection{Invariant shape anisotropy parameter $\kappa^2$}

\begin{figure*}[h]
\hspace*{-8mm} 
\subfigure[]{\includegraphics[width=8.9cm]{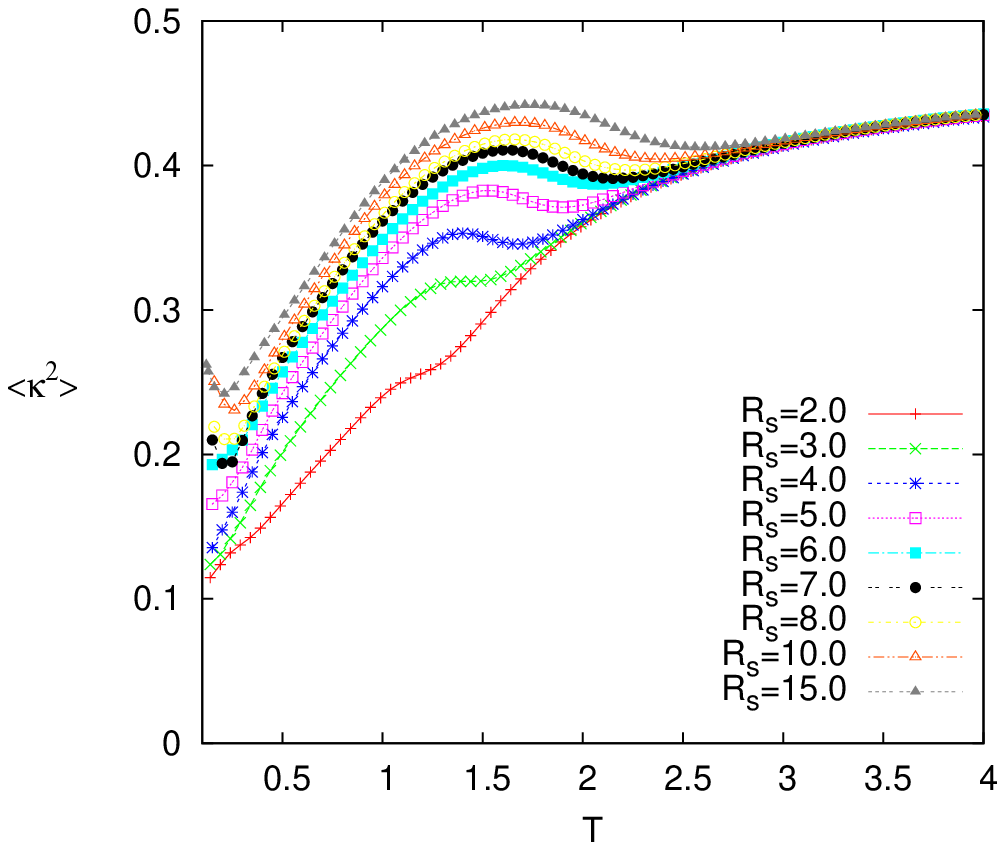}}
\subfigure[]{\includegraphics[width=8.9cm]{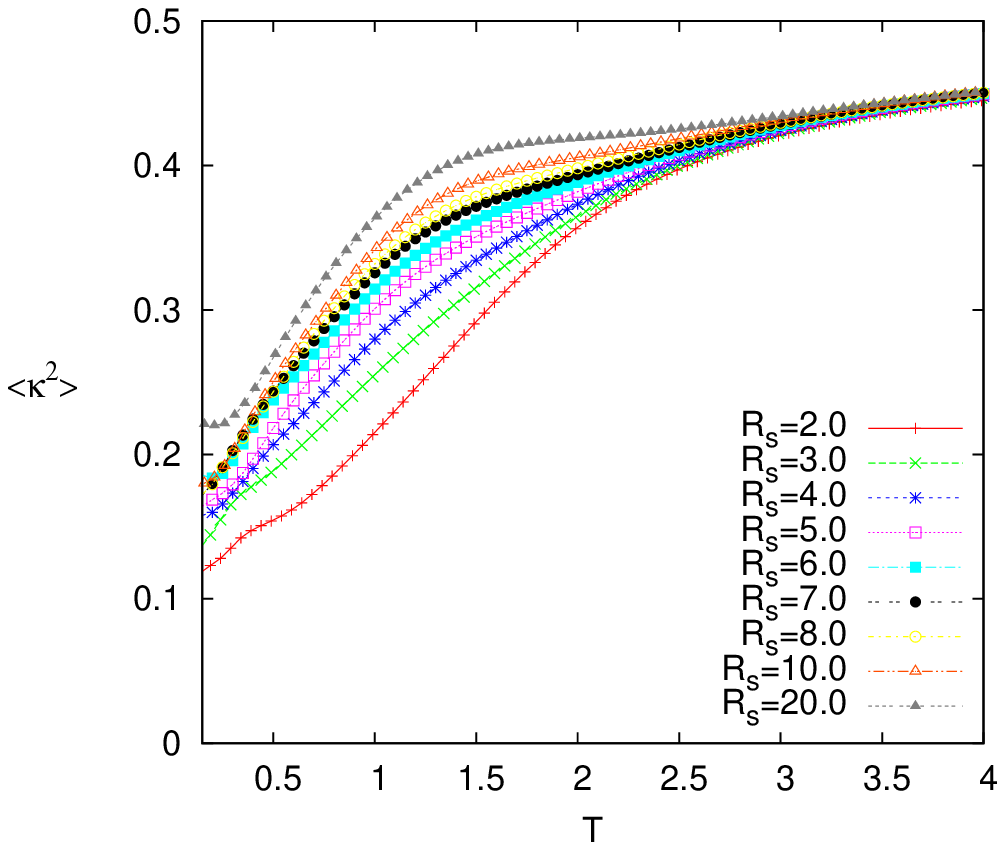}}
\hspace*{-8mm} 
\subfigure[]{\includegraphics[width=9.0cm]{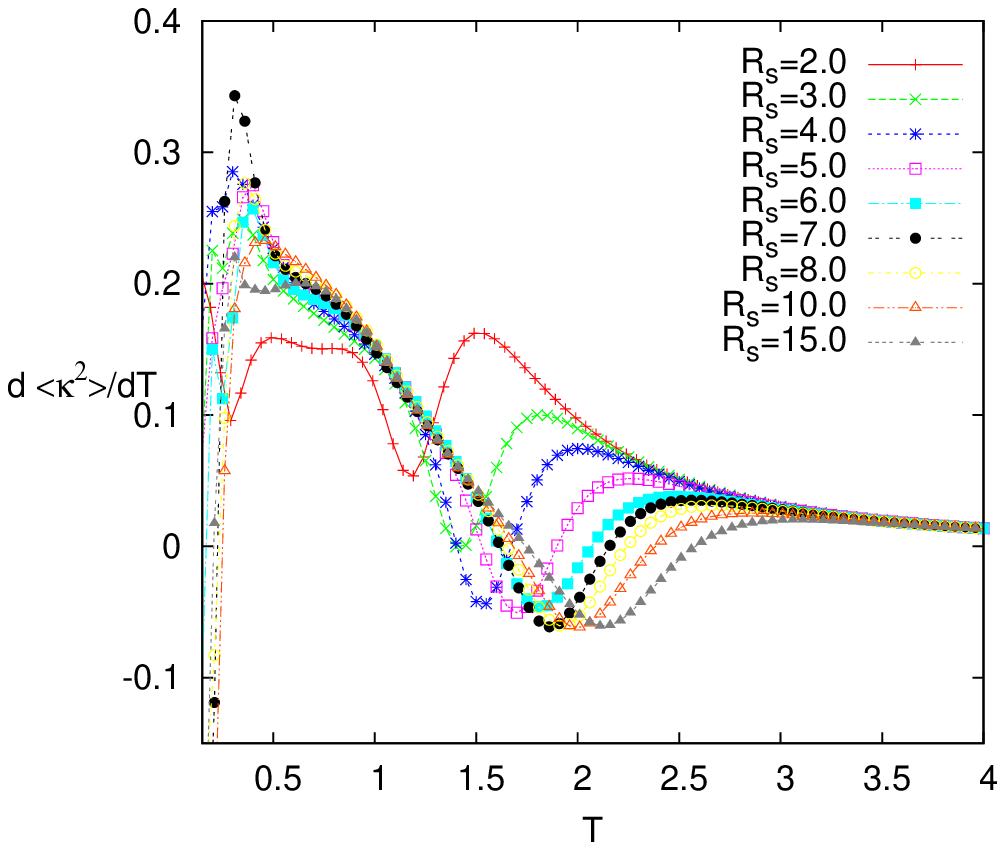}} 
\subfigure[]{\includegraphics[width=8.9cm]{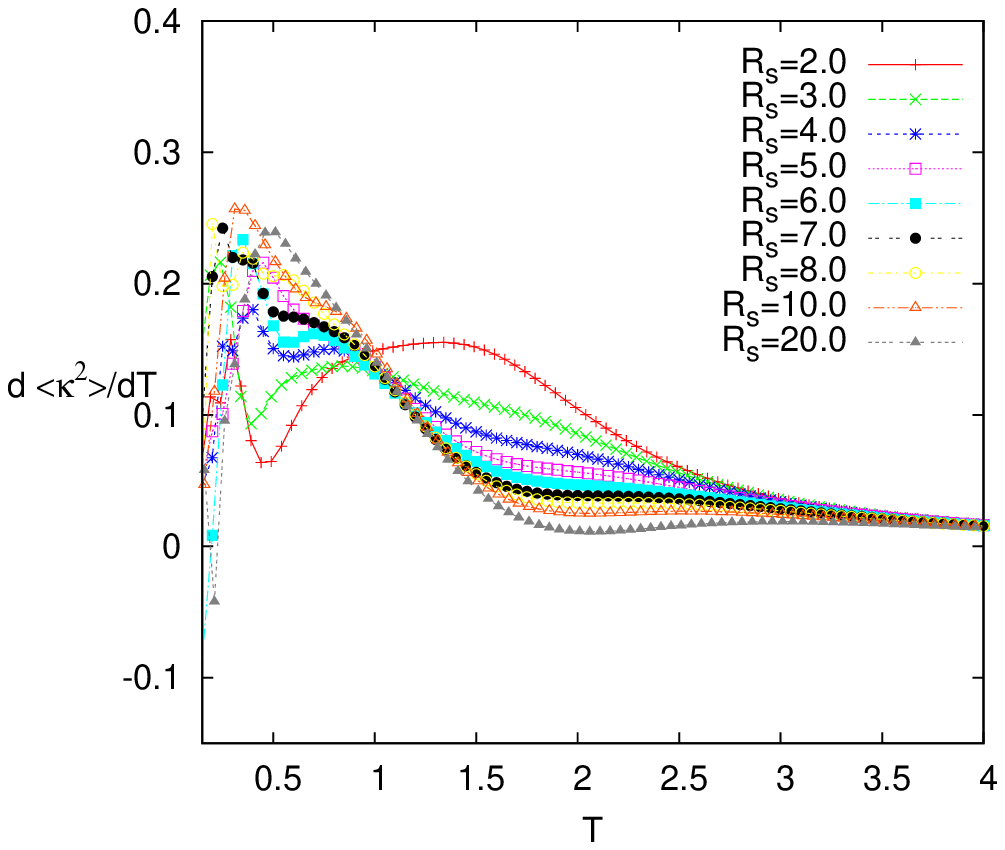}}
\caption{\label{fig:rgK} (Color online)  The canonical expectation value of the relative shape 
anisotropy parameter $\langle \kappa^2 \rangle $ for the (a) non-grafted and (b) end-grafted case,
and (c), (d) the corresponding temperature derivatives, for different sphere radii  $R{\rm_s}$
(polymer length $N=20$, surface attraction strength $\epsilon=1.0$).
} 
\end{figure*}

	
Also the relative shape anisotropy parameter $\langle \kappa^2 \rangle$ (the second invariant 
of the gyration tensor) presented in Fig.~\ref{fig:rgK} 
gives rich information and supports our findings derived from $\langle R_{\rm g}\rangle$.  
As discussed above, in the non-grafted case, the desorbed phase is divided into two regions 
which are called ``desorbed$_1$'' and ``desorbed$_2$''. The boundary between these two regions 
is emerging in the temperature derivative 
$d\langle \kappa^2 \rangle/dT$ 
displayed in Fig.~\ref{fig:rgK}(c)
(and also in $d\langle R_{\rm g}\rangle/dT$) as a second peak at high
temperatures, since the peaks are going to become invisible with increasing $R{\rm_s}$ values and 
also are smaller than the peaks at 
low temperatures. We have investigated the conformations in both regions in detail and concluded 
that the conformations in the ``desorbed$_1$''  phase are far away from the surface. On the other 
hand, the conformations in the ``desorbed$_2$'' phase are almost adsorbed to the sphere boundary, 
which indicates the influence of the surface on the desorbed phase.       
Additionally, the relative shape anisotropy parameter $\langle \kappa^2 \rangle$
clearly gives the phase boundaries at very low temperatures (below the freezing
transition at $T \approx 0.3$). The curves in Fig.~\ref{fig:rgK}(a) belonging to different 
$R{\rm_s}$ values are grouped at very low temperatures  into  different $\kappa^2$ 
values, indicating the boundaries from compact to two-layer phase, and from two-layer 
to monolayer phase in the phase diagrams.  

\begin{figure*}[h]
\hspace*{-8mm}
\subfigure[]{\includegraphics[width=8.9cm]{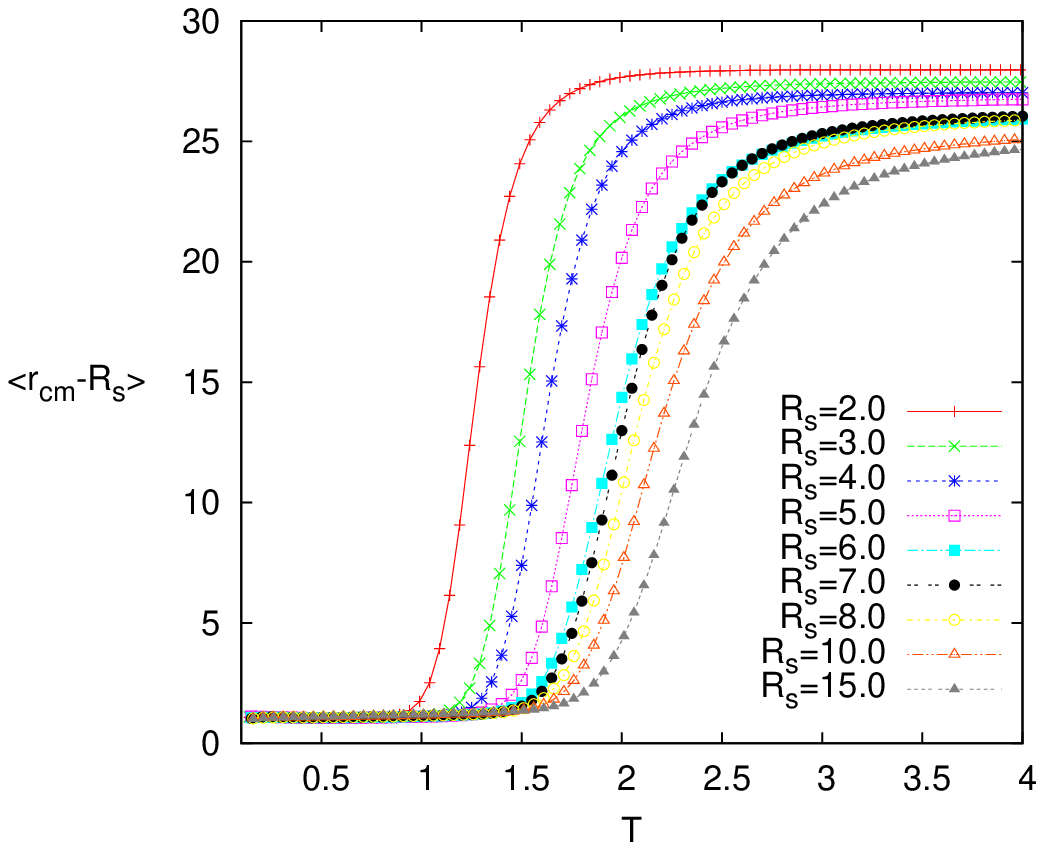}}
\subfigure[]{\includegraphics[width=8.9cm]{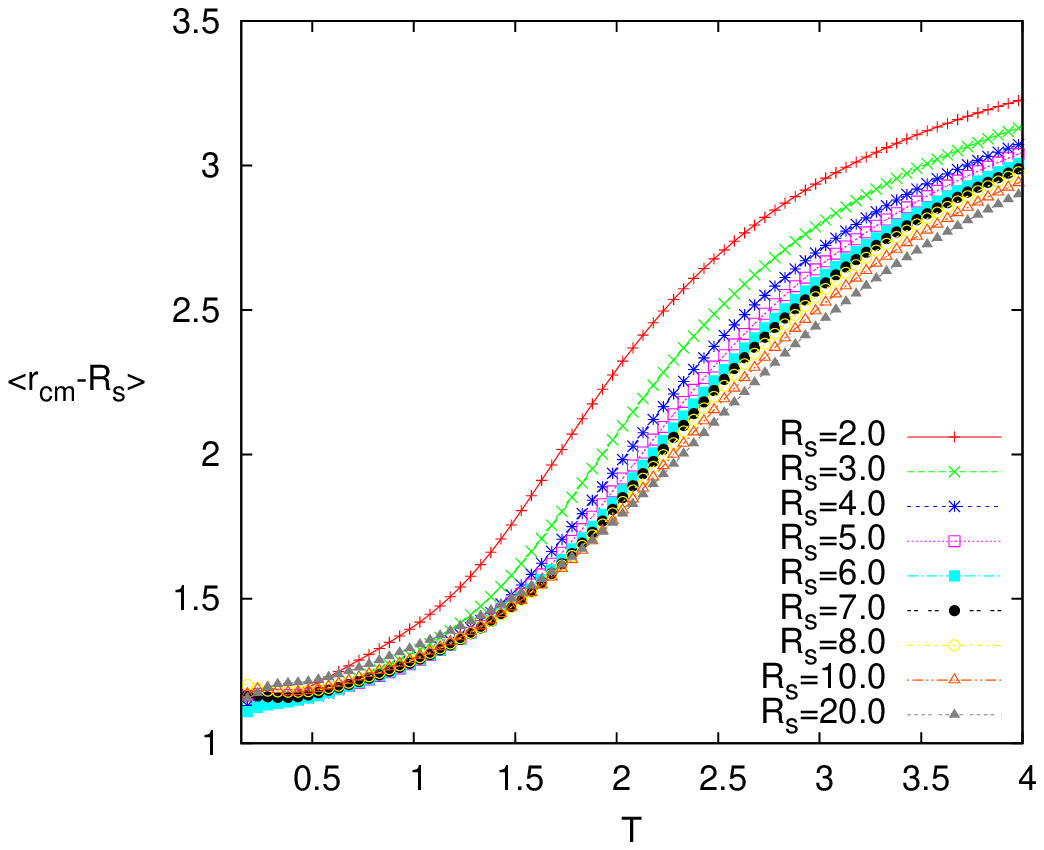}}
\caption{\label{fig:rcm} (Color online) The canonical expectation value of the distance of 
the center-of-mass of the polymer  $\langle r_{\rm cm} \rangle$ from the sphere surface  
for the (a) non-grafted and (b) end-grafted case 
(polymer length $N=20$, surface attraction strength $\epsilon=1.0$).
}
\end{figure*}

\begin{figure*}[h]
\hspace*{-8mm} 
\subfigure[]{\includegraphics[width=8.9cm]{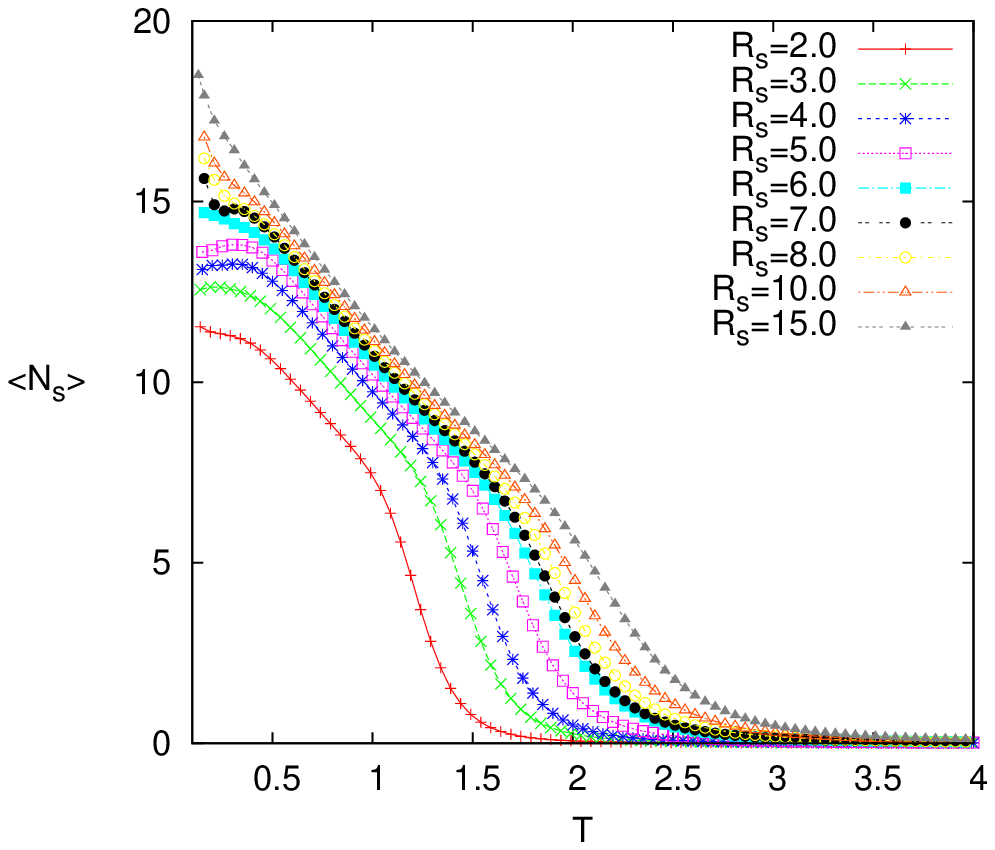}}
\subfigure[]{\includegraphics[width=8.9cm]{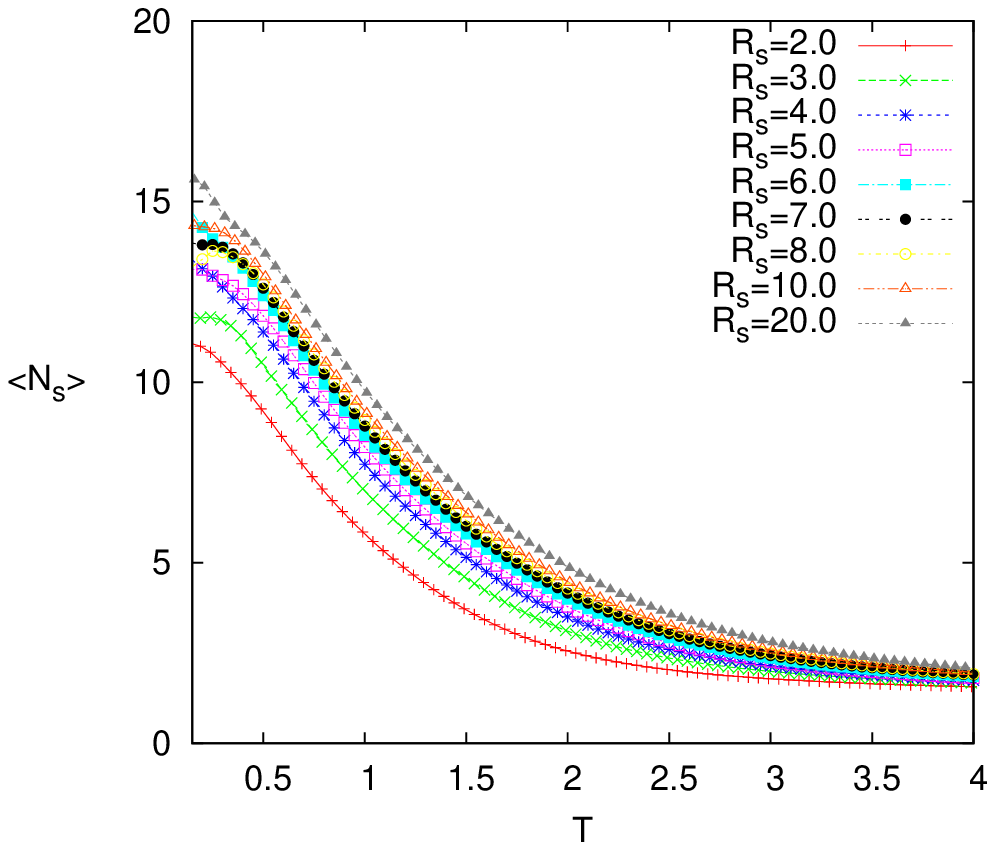}}
\caption{\label{fig:ns} (Color online) The canonical expectation value of the mean number of 
	docked monomers $\langle N_{\rm s} \rangle$ for the (a) non-grafted and (b) end-grafted case
(polymer length $N=20$, surface attraction strength $\epsilon=1.0$). 
}
\end{figure*}

\subsubsection{Center-of-mass distance $r_{\rm cm}$ and number of adsorbed monomers $N_{\rm s}$}

The adsorption transition can be best detected by the distance of the center-of-mass of the polymer
to the substrate $r_{\rm cm}-R_{\rm s}$ and by the number of adsorbed monomers $N_{\rm s}$, where a monomer 
is defined to be adsorbed onto the surface if $r_i-R_{\rm s} < 1.2$. The behavior of these two observables,
in particular the peaks in their temperature derivative, build the adsorption line in the phase diagrams.
Figures~\ref{fig:rcm}(a) and (b) give the distance
of the center-of-mass of the polymer to the sphere surface for the
non-grafted and end-grafted cases, respectively.
As can be seen in Fig.~\ref{fig:rcm}(a), for high temperatures the non-grafted polymer can 
move freely within the simulation space and the influence of the surface is minimal for large 
$R{\rm_s}$ values, whereas for small $R{\rm_s}$ the influence is mainly steric. Thus, the average 
center-of-mass distance of the polymer above the surface is nearly half of the simulation space. 
In contrast, at low temperatures the polymer favors surface contacts and the average center-of-mass
distance converges to the minimum location of the potential (cf.\ Fig.~\ref{fig:pot}).  
One can clearly detect a quite pronounced peak in its temperature derivative [see Supplementary Material] 
that divides the phase space into adsorbed and desorbed phases. Consistently with our discussion above,
the pronounced tendency of the polymer to make surface contacts can also be identified from the mean 
number of adsorbed monomers to the surface $\langle N_{\rm s} \rangle$ shown in Fig.~\ref{fig:ns} and the 
(negative) minima in its temperature derivative [see Supplementary Material]. They are in good agreement 
with the sharp peaks in the temperature derivative of the distance of the center-of-mass of the polymer 
which together draw the adsorption line in the phase diagram.  
By comparing the end-grafted with the non-grafted case, the main difference
is found at the adsorption transition: A crossover occurs from low temperature, 
where the polymer is adsorbed and the conformations of an end-grafted and a non-grafted
polymer are very similar, to high temperatures, where the non-grafted polymer approaches
the behavior of a polymer in bulk solution while that of an end-grafted polymer is 
always affected by the attractive sphere surface. Because of this effect, the adsorption 
transition for the end-grafted chain is much smoother which can be clearly seen in the     
$\langle r_{\rm cm} \rangle $ and $\langle N_s \rangle$ parameters in Figs.~\ref{fig:rcm}(b) and
\ref{fig:ns}(b).
In contrast, the adsorption of a non-grafted chain exhibits a first-order-like 
signature which is also clear from the same structural parameters. Because in the
non-grafted case these quantities change sharply as soon as the polymer desorbs, it 
leaves the influence of the surface field. An end-grafted polymer, on the other hand, cannot 
leave the surface field.

\begin{figure*}[t]
\hspace*{-8mm} 
\subfigure[]{\includegraphics[width=8.9cm]{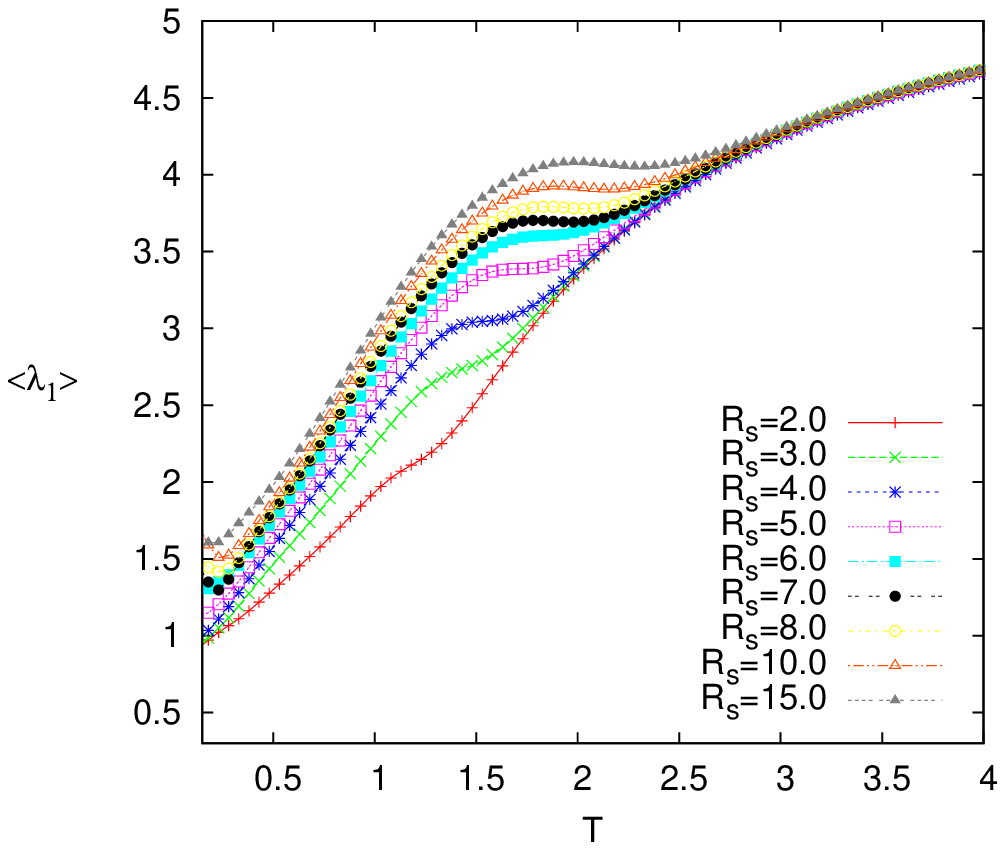}}
\subfigure[]{\includegraphics[width=8.9cm]{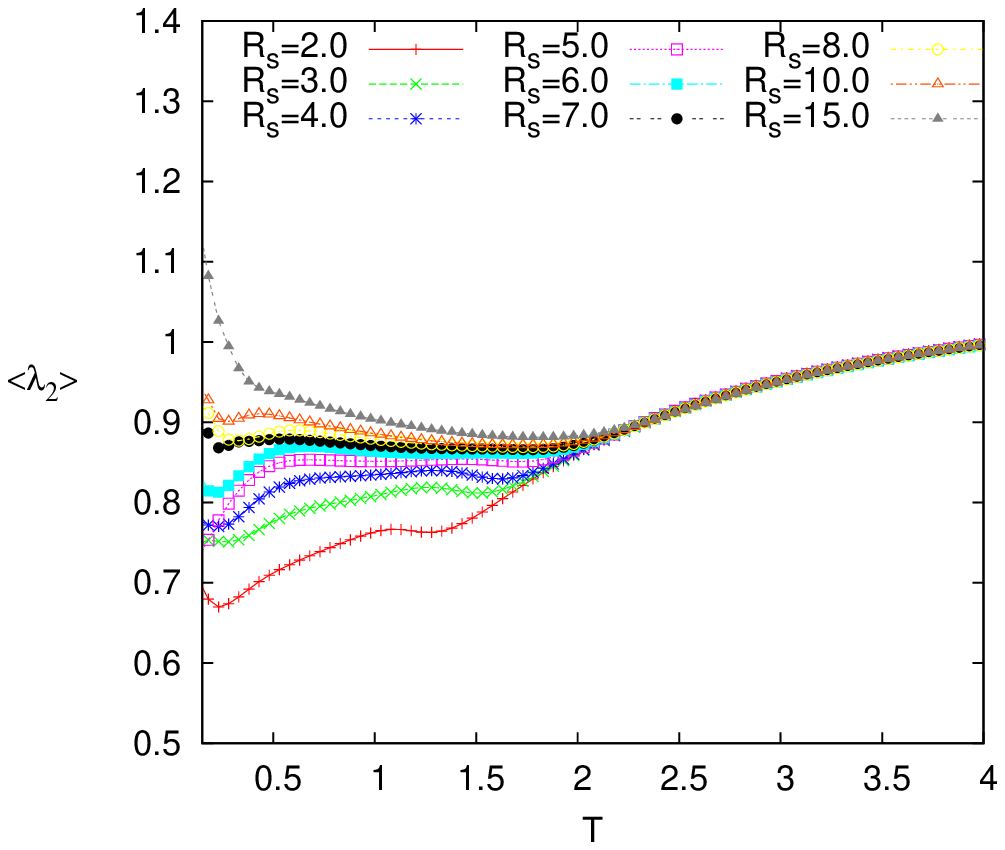}}
\hspace*{-8mm} 
\subfigure[]{\includegraphics[width=8.9cm]{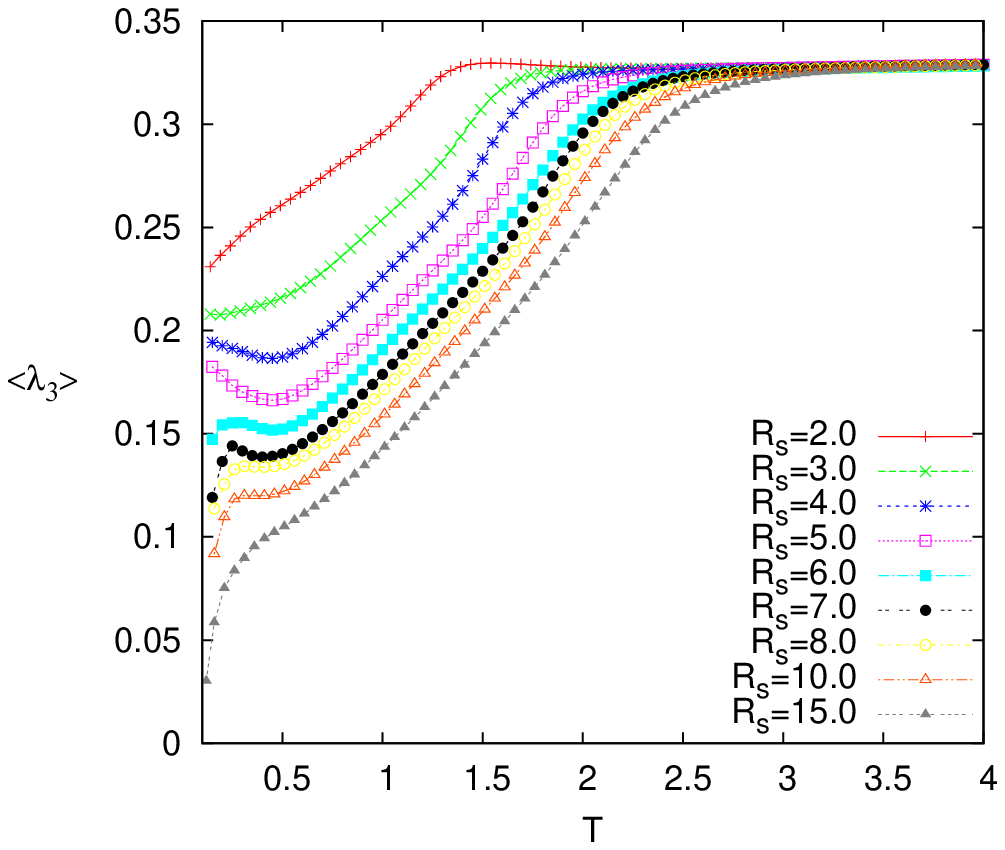}}
\subfigure[]{\includegraphics[width=8.9cm]{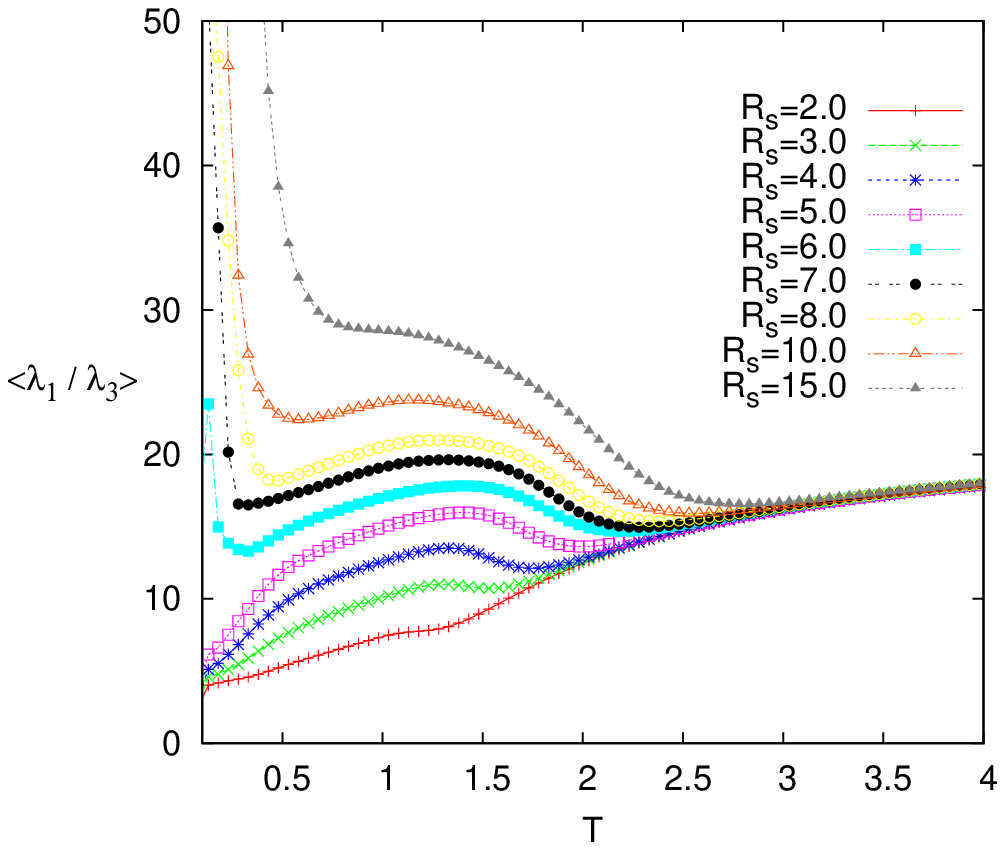}} 
\caption{\label{fig:eigen} (Color online)  The canonical expectation values
of the three eigenvalues $\langle \lambda_1 \rangle$, $\langle \lambda_2 \rangle$, 
and $\langle \lambda_3 \rangle$ of the gyration tensor and the ratio of the largest 
to the smallest eigenvalue $ \langle \lambda_1/\lambda_3\rangle$ for a non-grafted 
polymer in the presence of an attractive sphere with different radii $R_{\rm_s}$
(polymer length $N=20$, surface attraction strength $\epsilon=1.0$).
}
\end{figure*}


\subsubsection{Eigenvalues of the gyration tensor}

Finally, the eigenvalues of the gyration tensor which measure the extensions in the principle axis system 
are  extracted to  complement the picture.
In Figs.~\ref{fig:eigen}(a)-(c)  
they are displayed for different values of $R_{\rm s}$ for the non-grafted case. 
For high temperatures they are in good agreement
with the  results in our previous study ~\cite{Arkin2013a}, showing the 
same limit values of the three eigenvalues for random-coil structures,
and overall they all support our earlier findings described above. 
For low temperatures, the curves in Figs.~\ref{fig:eigen}(a)-(c) belonging
to different $R_{\rm s}$ values are also grouped into different 
$\langle \lambda_1 \rangle$, $\langle \lambda_2 \rangle$ and 
$\langle \lambda_3 \rangle$ values, indicating the boundaries (grey bands) 
in the phase diagrams which are detected from the other structural quantities.    
The most important
result deducible from the eigenvalues is that: The third eigenvalue of the gyration tensor 
$\langle \lambda_3 \rangle$ converges to small values which means that the extension in the 
third direction vanishes and the conformations are two-dimensional objects signaling the 
layering transition. To highlight this finding we show in Fig.~\ref{fig:eigen}(d) the ratio 
of the largest to the smallest eigenvalue
$ \langle \lambda_1 / \lambda_3 \rangle$.
For low temperatures below $T \approx 0.3$,
this ratio assumes relatively small values until  $R_{\rm s} \approx 6.0 $. Above 
$R_{\rm s} \approx 7.0 $ the 
ratio of the eigenvalues jumps to very much larger values, confirming
that the  layering transition occurs at this $R_{\rm s}$ value, which separates the
conformational space from planar conformations. This signal is also reflected in the 
corresponding temperature derivatives
 which are compiled in the Supplementary Material. 

\section{Conclusion}
In this paper, we have reported results from extensive multicanonical Monte Carlo computer 
simulations for investigating the full conformational behavior of a 
generic coarse-grained finite polymer chain near an attractive spherical surface.
In a systematic analysis, over a wide range of sphere radius $R_{\rm s}$ and temperature $T$,
we have constructed the entire phase diagrams for both non-grafted and end-grafted polymers.
For the identification of the conformational phases, we have examined several energetic 
and structural  observables and their fluctuations by canonical statistical analysis.  
The transition lines in the phase diagrams show the best match of all observables analyzed 
simultaneously in our study. In the thermodynamic limit of infinitely long chains the 
transitions are expected  to occur at sharp values of the parameters. For finite chains, 
on the other hand, the transition lines still vary with chain length $N$ and are not well 
defined because of broad peaks in the observables that also have small differences in between.
Therefore the locations of the phase boundaries should be considered as a rough guide.
However, even for the rather short chains considered here, 
we can clearly identify different phases which show distinguishing features, so that
a reasonable picture is obtained.
Most of the phases are believed to still persist for longer chains.
All our results obtained from the different observables are summarized in the phase diagrams
in the $R{\rm_s}-T$ plane which for a convenient overview are displayed at the beginning
of the results section in Fig.~\ref{fig:phase}(a) for non-grafted and in Fig.~\ref{fig:phase}(b) 
for end-grafted polymers, respectively.

It is clear that, for longer chains, the desorbed, globule and compact phases  will survive. 
Additionally, filmlike (monolayer) and semispherical conformations (two layer), as well as 
surface attached globular shapes will dominate the respective phases. On the other hand, as 
long as surface effects are as influential as volume effects the compact adsorbed conformations
differ noticeably for polymers with different but small lengths. But, for the majority of phases 
we find qualitative coincidence with a simple coarse-grained model.   

In this study, we kept the adsorption field constant (whereas we varied the adsorption strength
in another earlier study) and varied the radius of the nano-particles and observed qualitatively 
the described scenarios. As a result, our model system can be mapped in the considered parameter 
range to real systems considered in experiments. For example, based on early experimental 
results~\cite{Lesins1, Lesins2}, Feng and Ruckenstein~\cite{Ruckenstein} examined the adsorption
of a specific polyampholyte chain on a single spherical nano-particle with three different radii.
In this application, the charge density at the particle surface regulates the strength of the 
adsorption field (corresponding to our parameter $\epsilon$) and the polymer composition regulates
the location of the coil-globule transition (corresponding to $\epsilon_{\rm LJ}$).
Compared with experimental findings, computational studies of generic coarse-grained models have 
the advantage that different combinations of parameters can be varied over wide ranges. In this
way, a specific detailed system can be put into a broader context and a deeper understanding based on 
fundamental principles of statistical physics can be gained.     

\section*{ACKNOWLEDGMENTS}

We wish to thank Viktoria Blavatska and Bernd Abel for useful discussions.
H.A. acknowledges support by the Alexander von Humboldt Foundation
under the Experienced Researcher Fellowship Programme.
W.J. thanks the German Research Foundation (DFG) for support
under Grant Nos.\ JA483/24-3 and SFB/TRR 102 (Project B04). 
We also benefitted from the Marie Curie IRSES network DIONICOS under 
Contract No.\ PIRSES-GA-2013-612707 within the European Union 
Seventh Framework Programme.
The computer time for the Monte Carlo simulations was provided by
NIC, Forschungszentrum J\"{u}lich, under Grant No.\ hlz24, which we 
gratefully acknowledge.


\end{document}


\title{ Supplementary Material for: Polymer Adsorption on Curved Surfaces}
\author{Handan Arkin}
\email[E-mail: ]{Handan.Olgar@eng.ankara.edu.tr}
\affiliation{Institut f\"ur Theoretische Physik,
Universit\"at Leipzig, Postfach 100\,920, D-04009 Leipzig, Germany }
\affiliation{Department of Physics Engineering, Faculty of Engineering, Ankara University,
Tandogan, 06100 Ankara, Turkey}
\author{Wolfhard Janke$^{1}$}
\email[E-mail: ]{Wolfhard.Janke@itp.uni-leipzig.de}
\homepage[\\ Homepage: ]{http://www.physik.uni-leipzig.de/CQT.html}

\begin{abstract}

\end{abstract}

\maketitle

 Next to the  expectation  
values $\left\langle O\right\rangle$ of all structural quantities in  our main text, 
we also determined the fluctuations of these
structural quantities, as represented by the temperature derivative $d\langle O \rangle/dT = \left( \left\langle O E \right\rangle -\left\langle O\right\rangle \left\langle E\right\rangle \right) /T^2$.
We use generic units, in which $k_B=1$. The fluctuations of all structural quantities not discussed in the main text  are given in 
 the following figures:

\begin{figure*}[h]
\centering
\hspace*{-8mm}
\subfigure[]{\includegraphics[width=7.9cm]{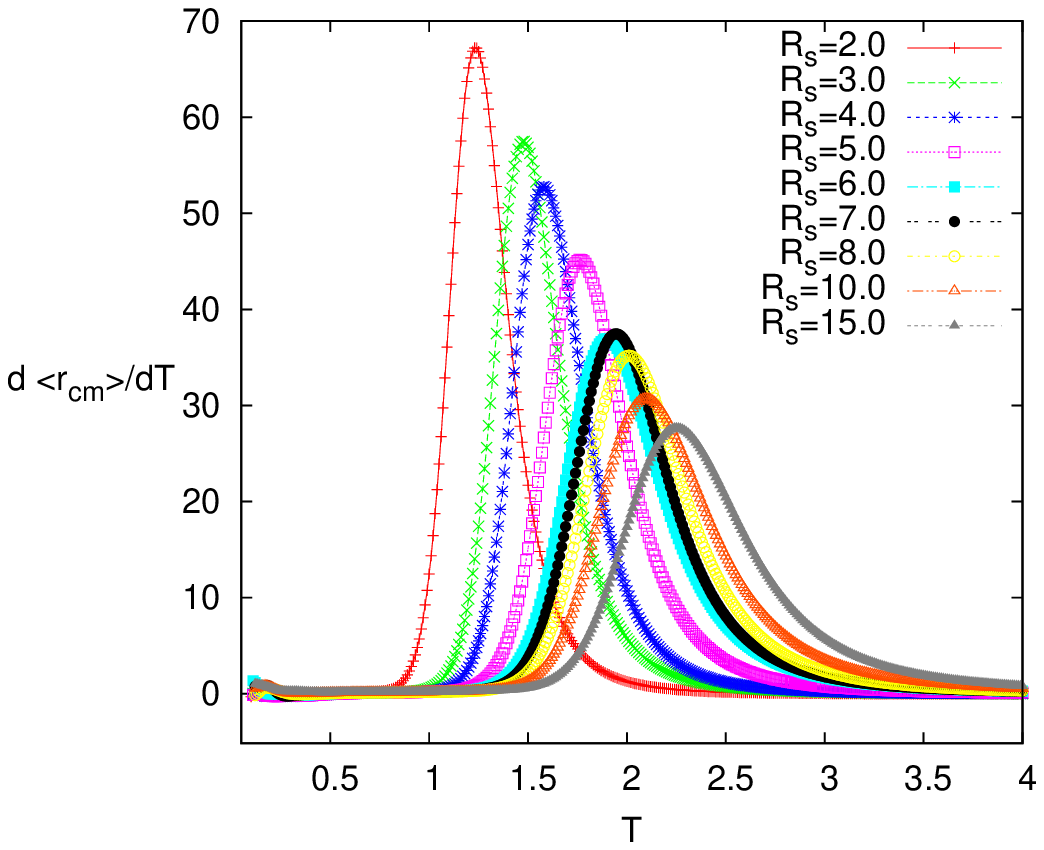}}
\subfigure[]{\includegraphics[width=7.9cm]{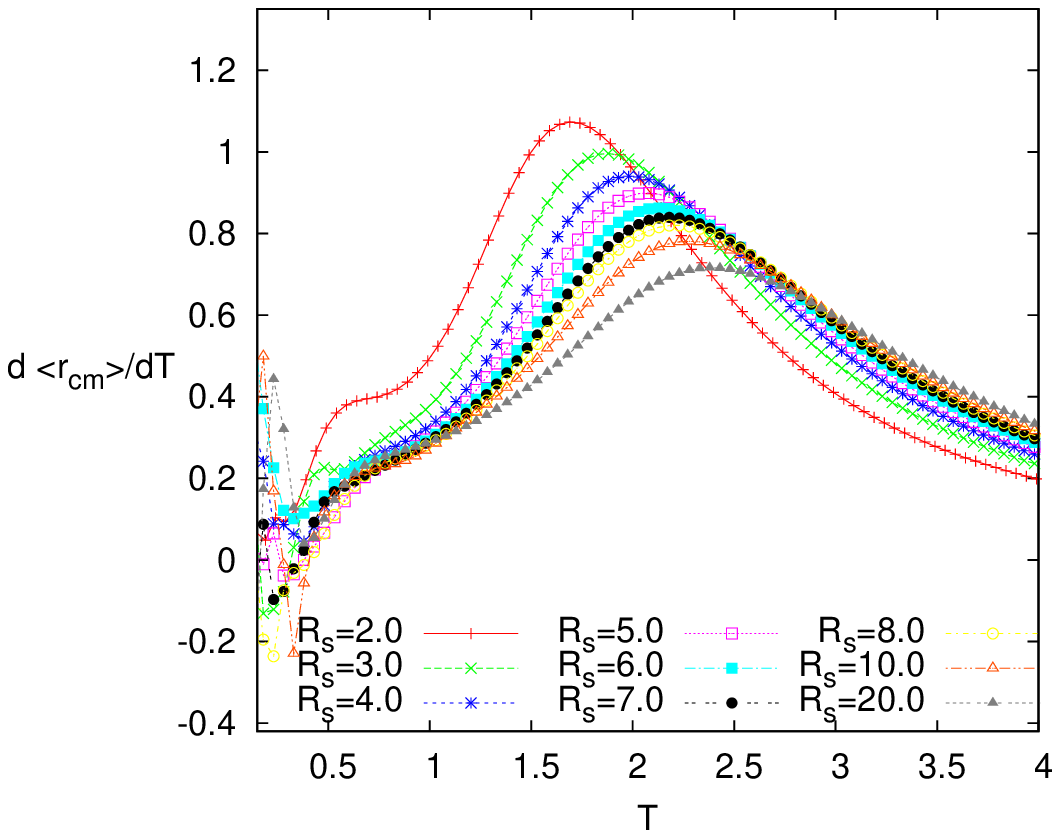}}
\caption{\label{fig:fluctuations} (Color online) The fluctuations of 
 the distance of the center-of-mass of the polymer from the sphere surface for (a) the non-grafted and (b) the end-grafted case.}
 \end{figure*}

\begin{figure*}[h]
\centering
\hspace*{-8mm}
\subfigure[]{\includegraphics[width=7.9cm]{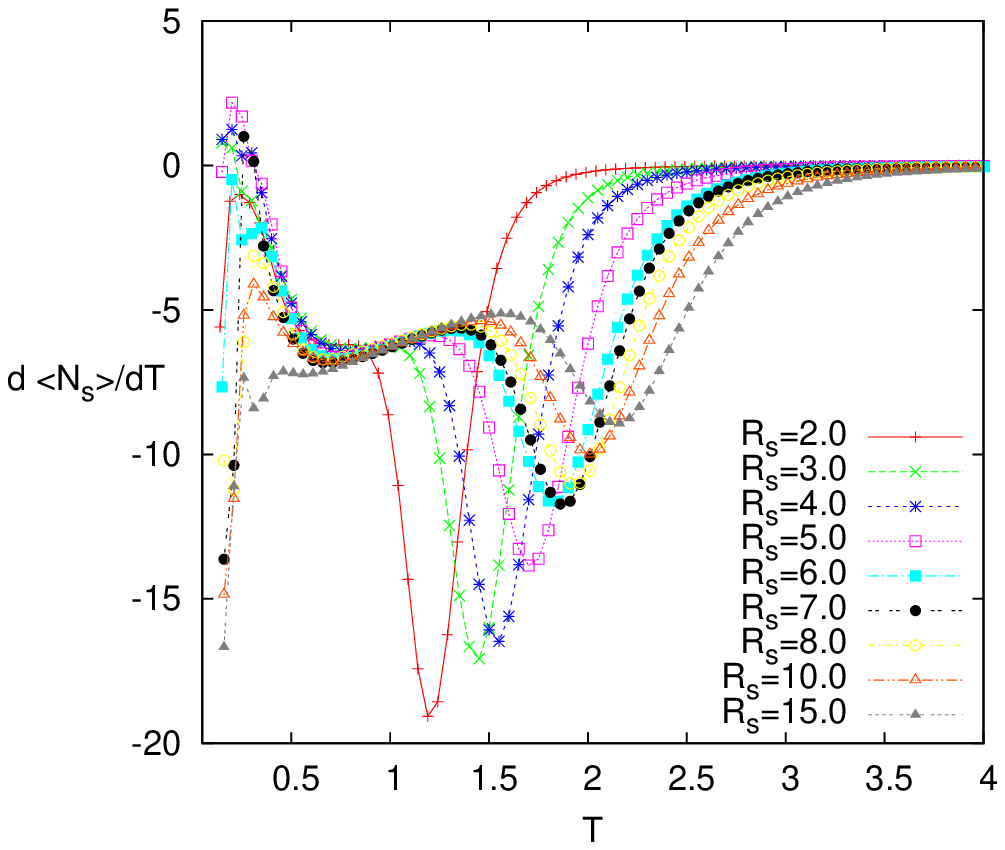}}
\subfigure[]{\includegraphics[width=7.9cm]{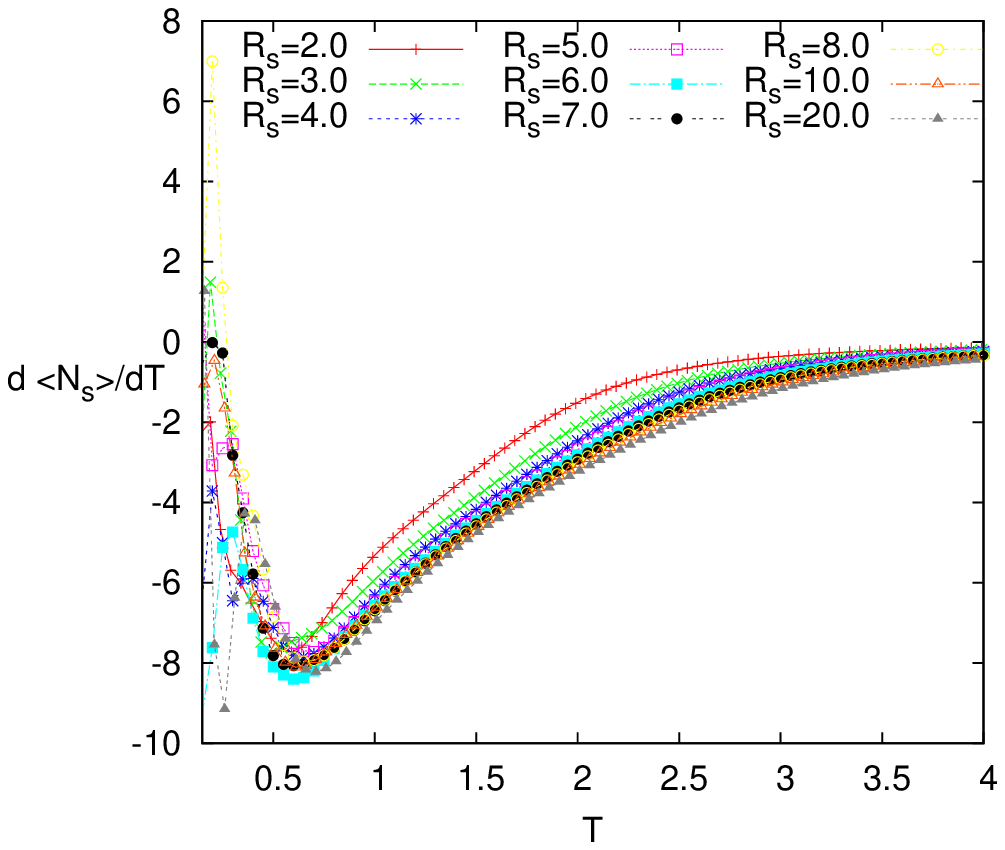}}
\caption{\label{fig:fluctuations} (Color online) The fluctuations 
of  the  number of 
adsorbed  monomers for (a) the  non-grafted  and (b) the  end-grafted case.}
\end{figure*}

\begin{figure*}[t]
\hspace*{-8mm}
\subfigure[]{\includegraphics[width=7.9cm]{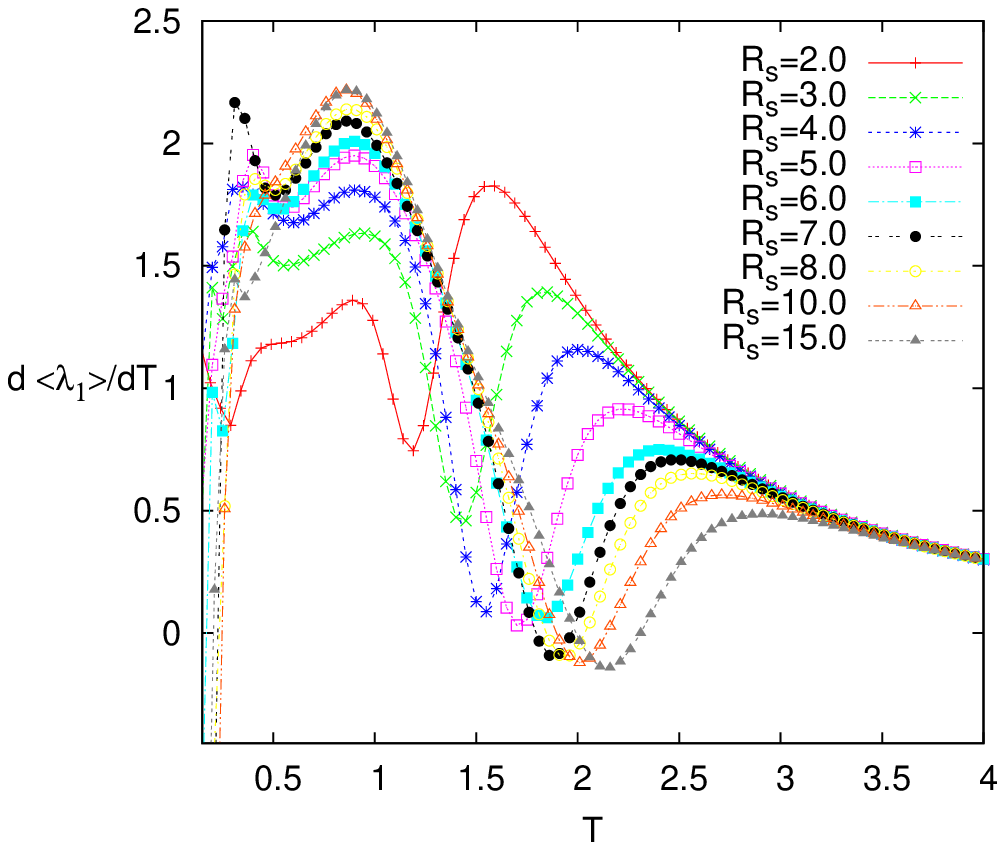}}
\subfigure[]{\includegraphics[width=7.9cm]{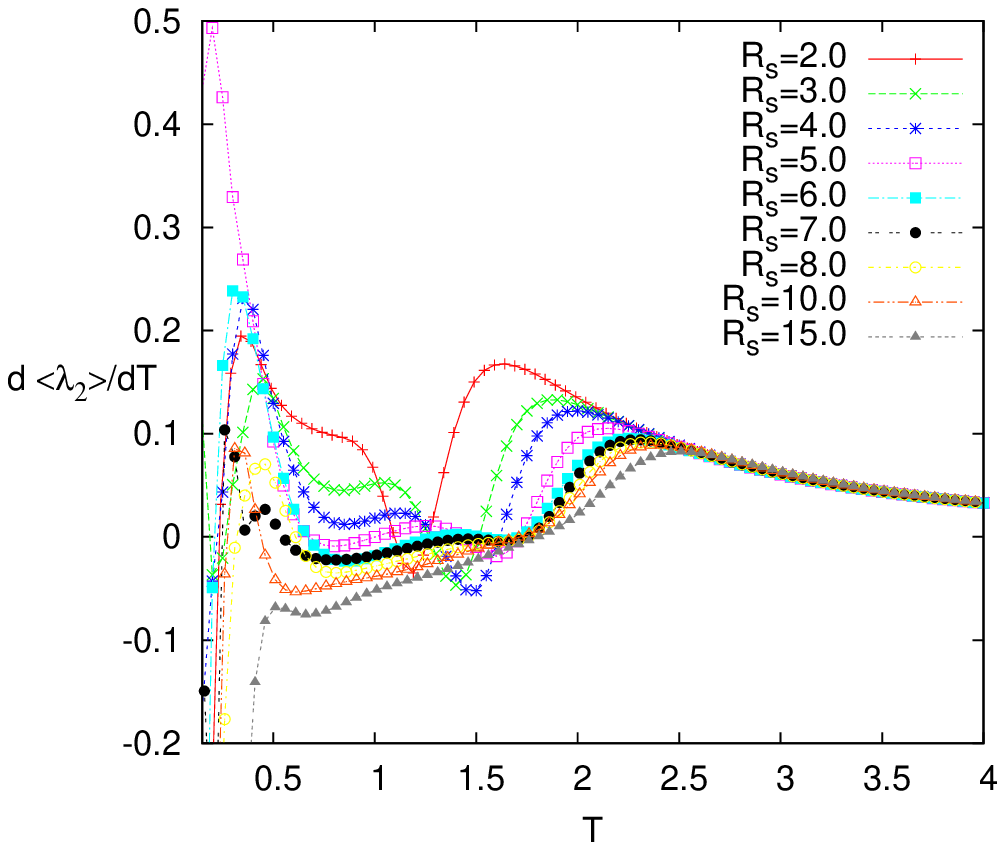}}
\subfigure[]{\includegraphics[width=7.9cm]{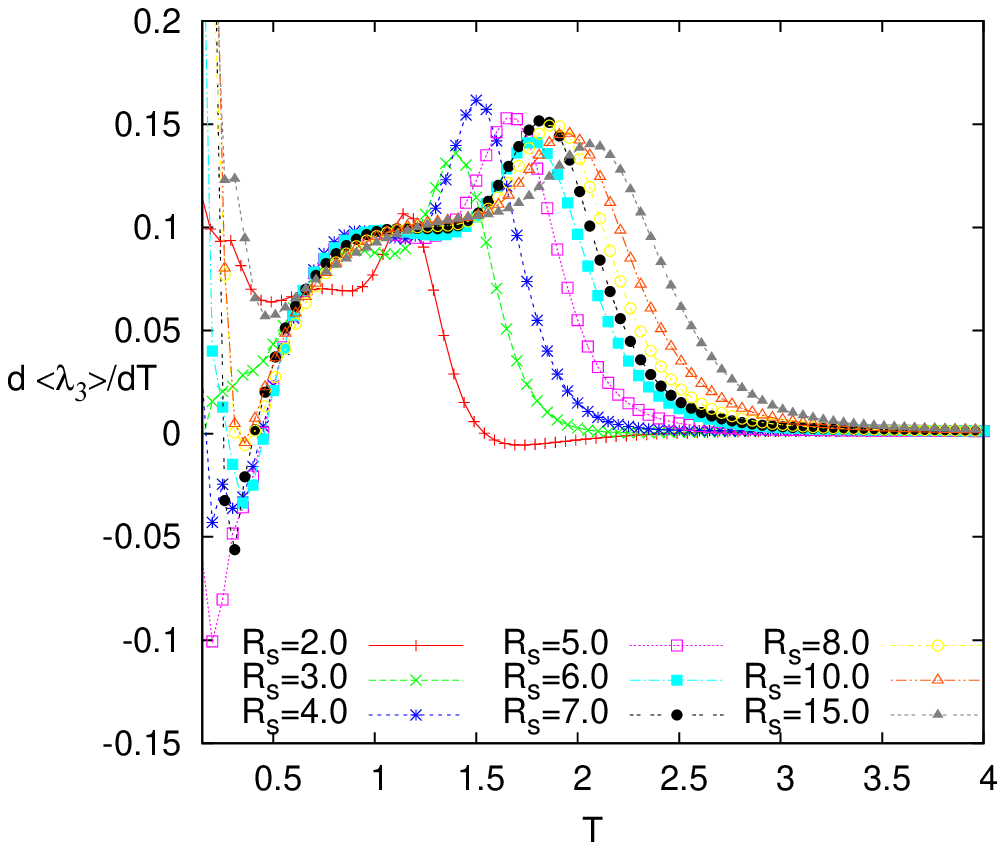}}
\subfigure[]{\includegraphics[width=7.9cm]{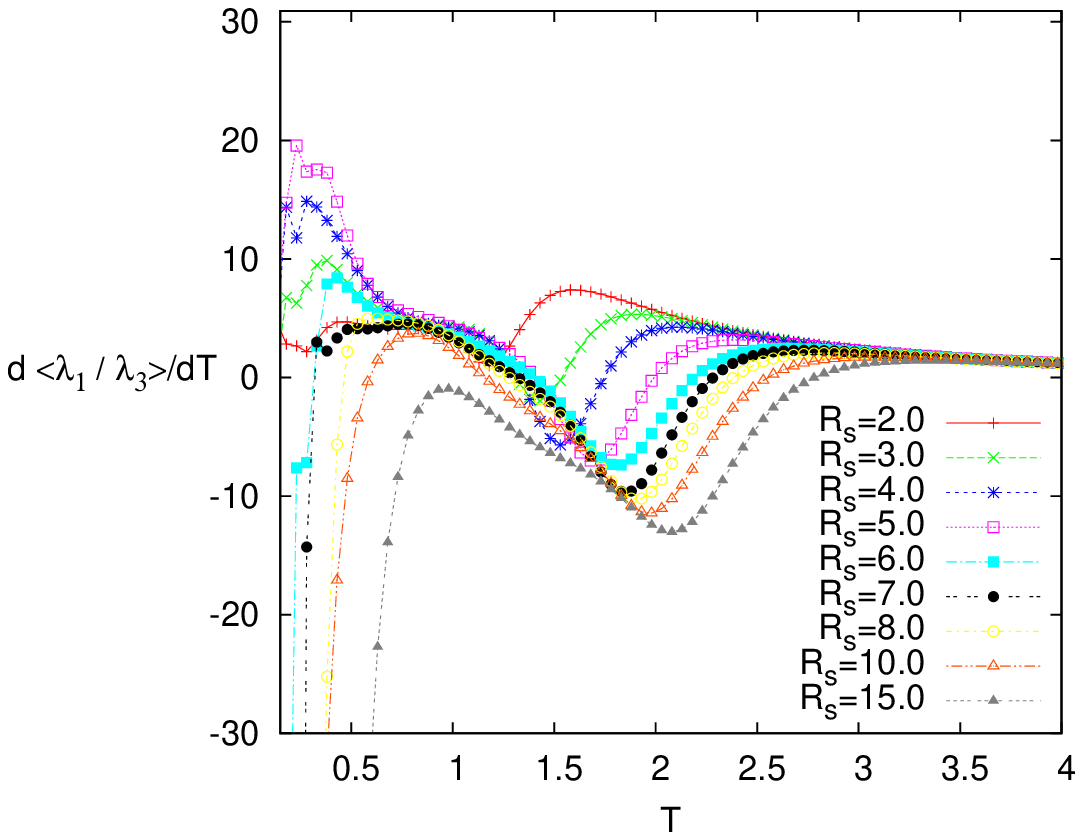} }
\caption{\label{fig:eigen} (Color online)  The fluctuations  of the three eigenvalues  (a) $\langle \lambda_1 \rangle$, (b) $\langle \lambda_2 \rangle$, (c)  $\langle \lambda_3 \rangle$ of the gyration tensor and (d) the fluctuations of the ratio of the largest eigenvalue to the smallest eigenvalue $ \langle \lambda_1/\lambda_3\rangle$  for different sphere radii $R{\rm_s}$ for the non-grafted case.  }
\end{figure*}